\documentclass[twocolumn,english,notitlepage,superscriptaddress,nofootinbib]{revtex4-1}
\usepackage[T1]{fontenc}
\usepackage[latin9]{inputenc}
\usepackage{float}
\usepackage{amsmath}
\usepackage{amssymb}
\usepackage{graphicx}
\usepackage{wasysym}

\makeatletter

\providecommand{\tabularnewline}{\\}

\usepackage{babel}

\newcommand{\affone}{Department of Physics and Astronomy, University College London, Gower Street, WC1E 6BT London, United Kingdom.}
\newcommand{\affthree}{School of Physics and Astronomy, University of Glasgow, Glasgow, G12 8QQ, United Kingdom.}
\newcommand{\afftwo}{ Vienna Center for Quantum Science and Technology, Faculty of Physics, University of Vienna, A-1090 Vienna, Austria.}

\makeatother

\usepackage{babel}
\begin{document}
\title{Coherent scattering 2D cooling in levitated cavity optomechanics}
\author{Marko Toro\v{s}}
\affiliation{\affone}
\affiliation{\affthree}
\author{Uro\v{s} Deli\'{c}}
\affiliation{\afftwo}
\author{Fagin Hales}
\affiliation{\affone}
\author{Tania S. Monteiro}
\affiliation{\affone}
\begin{abstract}
The strong light-matter optomechanical coupling offered by Coherent
Scattering (CS) set-ups have allowed the experimental realisation
of quantum ground state cavity cooling of the axial motion of a levitated
nanoparticle {[}U. Deli\'{c} et al., Science 367, 892 (2020){]}. An
appealing milestone is now quantum 2D cooling of the full in-plane
motion, in any direction in the transverse plane. By a simple adjustment
of the trap polarisation, one obtains two nearly equivalent modes,
with similar frequencies $\omega_{x}\sim\omega_{y}$ and optomechanical
couplings $g_{x}\simeq g_{y}$ -- in this experimental configuration
we identify an optimal trap ellipticity, nanosphere size and cavity
linewidth which allows for efficient 2D cooling. Moreover, we find
that 2D cooling to occupancies $n_{x}+n_{y}\lesssim1$ at moderate
vacuum ($10^{-6}$ mbar) is possible in a ``Goldilocks'' zone bounded
by $\sqrt{\kappa\Gamma/4}\lesssim g_{x},g_{y}\lesssim|\omega_{x}-\omega_{y}|\lesssim\kappa$,
where one balances the need to suppress dark modes whilst avoiding
far-detuning of either mode or low cooperativities, and $\kappa$
($\Gamma$) is the cavity decay rate (motional heating rate). With
strong-coupling regimes $g_{x},g_{y}\gtrsim\kappa$ in view one must
consider the genuine three-way hybridisation between $x$,{\normalsize{}{}{}{}{}$y$}
and the cavity light mode resulting in hybridized bright/dark modes.
Finally, we show that bright/dark modes in the levitated set-up have
a simple geometrical interpretation, related by rotations in the transverse
plane, with implications for directional sensing. 
\end{abstract}
\maketitle

\section{Introduction}

The coupling between light and matter has led to major milestones
in physics, from the Michelson-Morley experiment~\cite{michelson1887relative}
to the detection of gravitational waves by the LIGO collaboration~\cite{abbott2016observation}.
The basic scheme relies on the light acting as a probe -- offering
exceptional sensitivities -- which is now routinely done in state-of-the-art
optomechanical systems with high-quality mirrors. The latter are themselves
interesting systems and have led to the field of cavity optomechanics~\cite{aspelmeyer2014cavity}.
A mirror with a motional degree of freedom cooled to its ground state
is of particular interest as it becomes a quantum sensor and can thus
be used as a detector of weak forces and as a probe of the quantum-to-classical
transition~\cite{bose1997preparation}.

On the other hand quantum features of an object in all three spatial
dimensions -- with applications ranging from quantum foundations
to directional sensing -- can be explored using an optically levitated
nanoparticle~\cite{millen2020optomechanics,barker2010cavity,chang2010cavity,pender2012optomechanical,monteiro2013dynamics}.
Initial experimental efforts have been hindered by technical difficulties
of stable trapping in high vacuum~\cite{kiesel2013cavity,asenbaum2013cavity}
and several implementations have been considered such as hybrid tweezer-cavity
traps~\cite{romero2010toward,mestres2015cooling}, electro-optical
traps~\cite{millen2015cavity,fonseca2016nonlinear}, and trapping
in the near field of a photonic crystal~\cite{magrini2018near}.

Recently, a 3D coherent scattering (CS) setup was introduced to levitated
cavity optomechanics~\cite{delic2019cavity,windey2019cavity} using
methods adapted from atomic physics~\cite{vuletic2000laser,vuletic2001three,domokos2002collective,leibrandt2009cavity,hosseini2017cavity}.
In contrast to experiments that consider dispersive coupling, here
the cavity is driven solely by the dipole radiation of the optically
trapped silica particle. Due to the tight focus of the optical tweezer
this scheme yields unprecedentedly high optomechanical coupling rates,
which subsequently enabled ground-state cooling of the motion along
the cavity axis and thus opened the door to quantum levitated optomechanics~\cite{delic2020cooling}.

\begin{figure}[H]
\includegraphics[width=0.9\columnwidth]{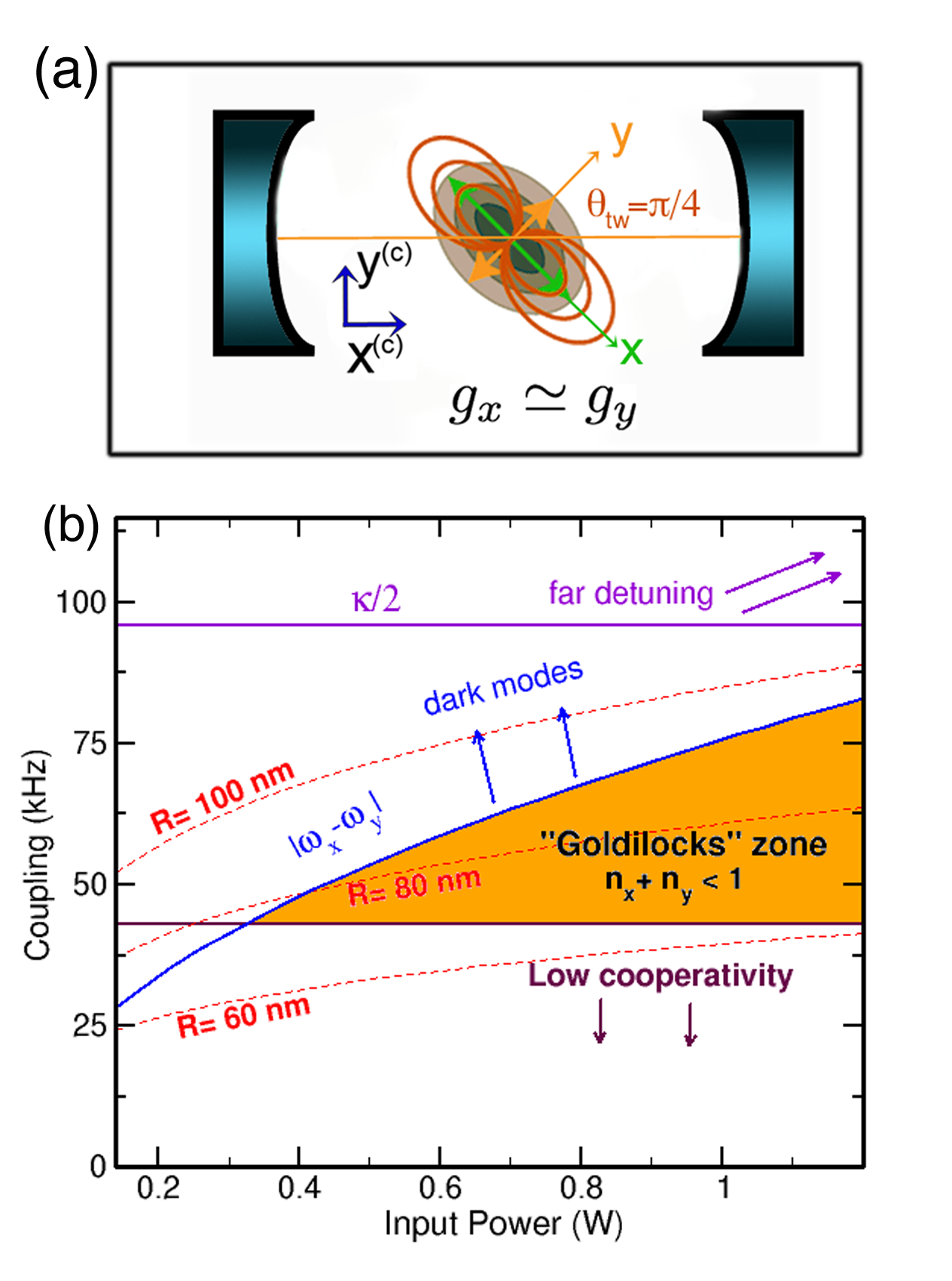} \caption{\textbf{(a)} Schematic of coherent-scattering experiments: an adjustment
of the tweezer polarisation ($\theta_{tw}=\pi/4,3\pi/4$) yields two
equivalently coupled $x,y$ mechanical modes, $|g_{x}|\simeq|g_{y}|\simeq g$.
\textbf{(b)} We find there is a ``Goldilocks'' region (orange) for
2D ground state cooling i.e. $n_{x}+n_{y}\lesssim1$, illustrated
for the set-up of ~\cite{delic2020cooling} but with $\theta_{tw}=\pi/4$.
The optimal region lies below the blue curve to avoid the formation
of decoupled dark modes (i.e., $|\omega_{x}-\omega_{y}|)\gtrsim g$),
is bounded from above by the constraint to avoid far-detuning ($\kappa\gtrsim|\omega_{x}-\omega_{y}|)$,
and from below by the regime of weak quantum cooperativities (i.e.,
$C=4g^{2}/(\kappa\gamma n_{B})\gtrsim1$, where $\gamma$ ($n_{B}$)
is the gas damping (mean thermal occupancy)). Red lines correspond
to different particle sizes and indicate $R\sim80$ nm is optimal.
\label{Fig1}}
\end{figure}

For the purpose of prolonging available free fall experiment times~\citep{hebestreit2018sensing},
an important future milestone for the coherent scattering setup is
the simultaneous ground state cooling of all three translational degrees
of freedom.

There is strong motivation, however, for investigating the cavity
cooling of 2D motions in the tweezer transverse plane ($x$-$y$ plane):
the frequencies are similar $\omega_{x}\approx\omega_{y}$ and for
suitable experimental parameters, $g_{x}\simeq g_{y}\equiv{g}=(g_{x}+g_{y})/2$,
so both may be strongly coupled to the light. In contrast, $g_{z}\sim0$
at the favourable configuration of trapping at the node of the cavity
that minimises optical heating, a key advantage of CS set-ups. In
addition, the frequency of the $z$-motion is typically $\omega_{z}\ll\omega_{x,y}$,
far from the optomechanical resonance. Finally, since the $z$ motion
can be cooled using feedback cooling~\cite{tebbenjohanns2020motional,magrini2020optimal}
strong 2D $x-y$ cavity cooling lays a path to 3D motional ground-state
cooling.

We exploit here recently developed theoretical expressions for the
full 3D coherent scattering problem ~\cite{torovs2020quantum}. However
although the numerics presented below are fully 3D, for analysis and
understanding we obtain and discuss the 2D $x-y$ problem. We optimize
2D cooling with respect to particle size, trap frequencies, tweezer
polarization orientation, as well as detuning between the tweezer
frequency and cavity resonance. For readily achieved experimental
pressures of $p=10^{-6}$ mbar we identify a ``Goldilocks'' region
$\sqrt{\kappa\Gamma/4}\lesssim g_{x},g_{y}\lesssim|\omega_{x}-\omega_{y}|\lesssim\kappa$,
where $\kappa$ ($\Gamma$) is the cavity decay rate (heating rate).
This set of requirements minimizes the formation of decoupled dark
modes and optimizes 2D cooling for $\vert\omega_{x}-\omega_{y}\vert\sim\kappa/2$
by using a particle of radius $\sim80\ \text{nm}$. While bright/dark
modes have been previously investigated in optomechanical systems~\cite{shkarin2014optically}
in the levitated system they have a geometric interpretation in terms
of the rotation of the $x$,$y$ axes of the oscillator, with potential
implications for directional sensing. The importance of non-degenerate
mechanical frequencies $\omega_{x}\neq\omega_{y}$ for successful
2D cooling is a well known fact in experiments with trapped ions and
atoms~\citep{neuhauser1978optical,leibrandt2009cavity}.

This work is organized in the following way. We start by reviewing
the coherent scattering setup and introducing the relevant experimental
parameters (Sec.~\ref{sec:Experimental-setup}). We then illustrate
how mechanical modes hybridize with the optical mode, resulting in
the formation of bright/dark modes and 3-way mixing. In particular,
we show how dark modes distort the relation between the displacement
and heterodyne spectra, making in general thermometry and sensing
non-trivial (Sec.~\ref{sec:Dark/Bright-modes-for}). 
In the central part we give a detailed analysis of 2D cooling and
discuss how to perform thermometry (Sec.~\ref{sec:analysis}) as
well as identify the best parameters for 2D cooling by numerically
scanning the experimental parameters (Sec.~\ref{sec:results}). We
conclude by laying down a path for 3D motional ground state cooling
in the levitated optomechanics -- in particular, how 2D cavity-optomechanical
cooling can be combined with feedback cooling to achieve the 3D motional
ground state of the optically levitated system (Sec.~\ref{sec:Discussion}).

\section{Experimental setup\label{sec:Experimental-setup}}

We consider the 3D coherent scattering setup illustrated in Fig.~\ref{Fig1}(a).
The nanoparticle is trapped in an optical tweezer and positioned inside
an optical cavity -- the cavity is driven entirely by the tweezer
light scattered off the nanoparticle, namely, coherent scattering
with a pattern shown in Fig.~\ref{Fig1}(a). Such a scheme offers
unique versatility with respect to the customary cavity optomechanical
system, since the nanoparticle can be placed at any point inside the
cavity by displacing the tweezer trap. Here we will consider the case
when the nanoparticle is close to a \emph{cavity} \emph{node} $x_{0}^{(\text{c})}\sim\lambda/4$
($\lambda$: laser wavelength), where the strongest coupling to the
nanoparticle $x$ and $y$ motions is achieved. In addition, deleterious
effects of cavity photon scattering and recoil heating are minimal.

Linearisation of the effective potentials in the CS set-ups ~\cite{delic2019cavity,windey2019cavity}
has shown that : 
\begin{equation}
g_{x}=E_{d}k\text{sin}(\theta_{\text{tw}})x_{\text{zpf}},\qquad g_{y}=E_{d}k\text{cos}(\theta_{\text{tw}})y_{\text{zpf}}\label{eq:couplings}
\end{equation}
where $E_{d}=\frac{\alpha\epsilon_{c}\epsilon_{\text{tw}}\text{sin}(\theta_{\text{tw}})}{2\hbar}$
is the driving amplitude of the cavity, $\alpha=3\epsilon_{0}V_{s}\frac{\epsilon_{R}-1}{\epsilon_{R}+2}$
($\epsilon_{0}$ is the permittivity of free space, $V_{s}$ is the
volume of the nanoparticle, $\epsilon_{R}$ is the relative dielectric
permittivity,), $\epsilon_{c}=\sqrt{\frac{\hbar\omega_{c}}{2\epsilon_{0}V_{c}}}$
($\omega_{c}$ is the cavity frequency, $V_{c}$ is the cavity volume),
$k=\omega_{c}/c$, and $\epsilon_{\text{tw}}=\sqrt{\frac{4P_{\text{tw}}}{w_{x}w_{y}\pi\epsilon_{0}c}}$
($P_{\text{tw}}$ is the tweezer power, and $w_{x}$, $w_{y}$ is
the waist of the Gaussian beam along the $x$ or $y$-axis respectively).

The angle $\theta_{\text{tw}}$ between the tweezer polarization axis
($y$-axis) and the cavity symmetry axis ($x^{\text{(c)}}$-axis)
can be arbitrarily set to tune coupling rates $g_{x}$ and $g_{y}$.
Motional 1D ground state cooling (of a single mechanical degree of
freedom) along $x^{\text{(c)}}$ has been recently achieved by setting
$\theta_{\text{tw}}\sim\pi/2$. In this case the\emph{ tweezer-based}
coordinates $(x,y)$ and the \emph{cavity-based} coordinates $(x^{\text{(c)}},y^{\text{(c)}})$
identify the same point in the 2D plane orthogonal to the tweezer
symmetry axis, $z$. However, one obtains $g_{x}\approx g_{y}$ for
$\theta_{tw}=\pi/4$ and that is the regime we consider for 2D cooling.

\section{Bright/Dark modes \label{sec:Dark/Bright-modes-for}}

Avoided crossings are ubiquitous in physics. For example, two classical
(or quantum) modes, say $\hat{x}$ and $\hat{y}$, approaching an
energy degeneracy are universally described by a Hamiltonian represented
in terms of Pauli matrices: 
\begin{equation}
\frac{\hat{V}_{\text{\text{int}}}}{\hbar}=\frac{1}{2}\left[\begin{array}{cc}
\hat{x} & \hat{y}\end{array}\right]\left[(\omega_{x}-\omega_{y})\hat{\sigma}_{z}+2g\hat{\sigma}_{x}\right]\left[\begin{array}{c}
\hat{x}\\
\hat{y}
\end{array}\right],
\end{equation}
where $g$ is the coupling. At the degeneracy, $\omega_{x}=\omega_{y}$,
the normal modes of the system correspond to the eigenmodes of $\hat{\sigma}_{x}$,
thus have the maximally hybridized form $\hat{x}\pm\hat{y}$. The
corresponding frequencies are perturbed by $\pm g$.

In regimes of negligible-dissipation $g\gg\kappa,\gamma$, the usual
optomechanical interaction corresponds to a two-mode avoided crossing.
The two-mode crossing was demonstrated experimentally in optomechanics~\cite{groblacher2009observation}
where it was observed that the cavity light mode and mechanical modes
were hybridized, with the characteristic $2g$ splitting, and a more
recent study investigated levitated nanoparticles~\cite{de2020strong}.

In the present case, we have a {\em three-mode} avoided crossing.
These are less frequently encountered, but may be discussed in a similar
way. In our present case, two mechanical modes, $\hat{x}$ and $\hat{y}$,
are coupled to the optical mode, $\hat{Z}_{L}$, according to the
usual position-position form (see Sec.~\ref{sec:analysis} for more
details): 
\begin{equation}
\frac{\hat{V}_{\text{\text{int}}}}{\hbar}=g_{x}\hat{x}\hat{Z}_{L}+g_{y}\hat{y}\hat{Z}_{L},\label{eq:2dOpto}
\end{equation}
Representing the modes as a vector, $[\hat{x}\,\hat{Z}_{L}\,\hat{y}]^{\top}$
and Eq.~\eqref{eq:2dOpto} in matrix form we write:

\begin{alignat}{1}
\frac{\hat{V}_{\text{\text{int}}}}{\hbar}=\frac{1}{4}\left[\begin{array}{ccc}
\hat{x} & \hat{Z}_{L} & \hat{y}\end{array}\right]\left[\begin{array}{ccc}
\omega_{x} & 2g_{x} & 0\\
2g_{x} & -\Delta & 2g_{y}\\
0 & 2g_{y} & \omega_{y}
\end{array}\right]\left[\begin{array}{c}
\hat{x}\\
\hat{Z}_{L}\\
\hat{y}
\end{array}\right]\label{eq:Sxmatrix}
\end{alignat}
where we have also included the two mechanical frequencies, $\omega_{x}$,
$\omega_{y}$, and the detuning, $-\Delta$.

We first consider equal couplings $g_{x}=g_{y}\equiv{g}$, and set
$-\Delta=(\omega_{x}+\omega_{y})/2$. Neglecting the term $\frac{1}{4}\frac{\omega_{x}+\omega_{y}}{2}\mathbb{I}$,
where $\mathbb{I}$ is the identity, we write: 
\begin{alignat}{1}
\frac{\hat{V}_{\text{\text{int}}}}{\hbar}=\frac{1}{4}\left[\begin{array}{ccc}
\hat{x} & \hat{Z}_{L} & \hat{y}\end{array}\right]\left[(\omega_{x}-\omega_{y})\hat{S}_{z}+2\sqrt{2}{g}\hat{S}_{x}\right]\left[\begin{array}{c}
\hat{x}\\
\hat{Z}_{L}\\
\hat{y}
\end{array}\right],\label{eq:simplified}
\end{alignat}
where now $\hat{S}_{x},\hat{S}_{z}$ are spin 1 matrices (divided
by $\hbar$). The associated anticrossing has an enhanced width of
$2\sqrt{2}g$.

\begin{figure}[ht]
\centering{}\includegraphics[width=0.72\linewidth]{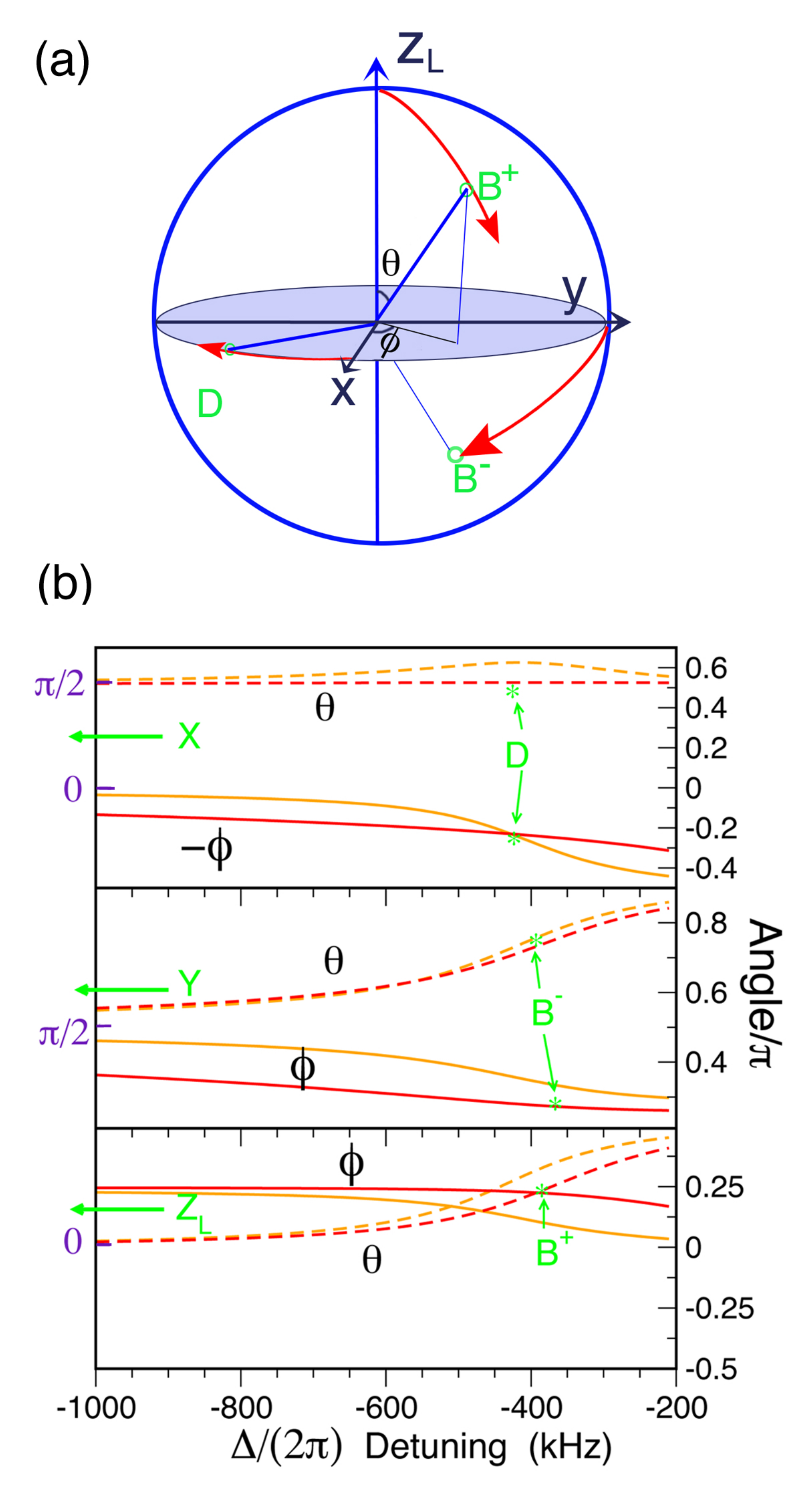}\caption{Trajectories of the eigenmodes of the 2D system, for $g_{x}=g_{y}$,
as a function of detuning $\Delta$. \textbf{(a)} Represents the trajectories
for the ideal $\omega_{x}=\omega_{y}$ case on a Bloch sphere, where
the vector of unit length $\hat{n}=[\hat{x},\hat{y},\hat{Z}_{L}]\protect\doteq[\theta,\phi]$
represents a mode at a given $\Delta$. The dark mode rotates along
the `equator' from $\phi=0$ at $\Delta=-\infty$, through an angle
$\phi=-\pi/4$ at $\Delta\to-(\omega_{x}+\omega_{y})/2$, thus evolves
from the $\hat{x}$ mode to $\hat{D}$. The `bright' modes rotate
from $\hat{Z}_{L}$ ($\hat{y}$) to $\hat{B}^{+}$ ($\hat{B}^{-}$).
\textbf{(b)} Shows the $\theta,\phi$ coordinates of the trajectories
for the three eigenmodes calculated from classical equations of motion, for realistic parameters:
either the elliptical trap (with $\delta\omega=|\omega_{x}-\omega_{y}|/((\omega_{x}+\omega_{y})/2)\simeq0.16$
of recent experiments ~\citep{delic2020cooling} (orange lines),
or a more circular trap (with $\delta\omega=|\omega_{x}-\omega_{y}|/((\omega_{x}+\omega_{y})/2)\simeq0.04$
(red lines). The near circular trap follows the idealised dark/bright
modes illustrated in (a), while for the elliptical trap the `dark'
mode still rotates by $\phi\simeq-\pi/4$ but does not remain on the
equator, thus mixes with the optical mode. The eigenmodes tend to
uncoupled $\hat{x},\hat{y},\hat{Z}_{L}$ modes at $\Delta\to-\infty$
(green arrows on the left) but hybridize into bright/dark modes $\hat{D},\hat{B}^{\pm}$,
as $\Delta\to-(\omega_{x}+\omega_{y})/2\sim-2\pi\times400\,\text{kHz}$
(green star symbols `{*}'). \label{spins}}
\end{figure}

For the degenerate case, $\omega_{x}=\omega_{y}$, the eigenmodes
are simply the textbook eigenvectors of the $\hat{S}_{x}$ matrix:
in that case, there are two \emph{three-way} hybridised ``bright''
eigenmodes ($\hat{B}^{\pm}$) with eigenvalues $\pm\sqrt{2}g$, and
a \emph{two-way} hybridised ``dark'' eigenmode ($\hat{D}$) with
eigenvalue zero: 
\begin{eqnarray}
\hat{B}^{\pm} & = & \frac{1}{2}[\hat{x}+\hat{y}\pm\sqrt{2}\hat{Z}_{L}],\\
\hat{D} & = & \frac{1}{\sqrt{2}}[\hat{x}-\hat{y}].\label{eq:dark}
\end{eqnarray}
While the pedagogic Eq.~\eqref{eq:simplified} was not used to compute
the realistic system eigenmodes, it illustrates the significance of
lifting the frequency degeneracy: $\omega_{x}\neq\omega_{y}$ introduces
a $\hat{S}_{z}$ component that mixes the bright dark modes $\hat{D},\hat{B}^{\pm}$.
The case $g_{x}\neq g_{y}$ (but $\omega_{x}\sim\omega_{y})$ does
not eliminate the bright-dark mode structure: it simply alters the
dark eigenvectors at the centre of the crossing to 
\begin{equation}
\hat{D}=\frac{1}{\epsilon}[g_{y}\hat{x}-g_{x}\hat{y}]\label{eq:gxneqgy}
\end{equation}
(which is still a dark mode with eigenvalue 0) while the bright modes
are $\hat{B}^{\pm}=\frac{1}{\sqrt{2}}[\frac{g_{x}}{\epsilon}\hat{x}+\frac{g_{y}}{\epsilon}\hat{y}\pm\hat{Z}_{L}]$
(with eigenvalues $\epsilon=\pm\sqrt{g_{x}^{2}+g_{y}^{2}}$). However,
we see that as $g_{y}\to0$, $\hat{D}\to\hat{y}$. This is the quasi-1D
dynamics analysed in \cite{torovs2020quantum} (using $\theta_{tw}=85$
degrees). In this limit, the $\hat{y}$ mode is ``dark'' simply
because it is very weakly coupled.

The true eigenmodes of the coherent scattering set-up for arbitrary
$\Delta$, were computed numerically from the equations of motion
of the system. The Hamiltonian for the reduced 2D case can be put
in the following form: 
\begin{alignat}{1}
\frac{\hat{H}}{\hbar}= & -\frac{\Delta}{4}(\hat{Z}_{L}^{2}+\hat{P}_{L}^{2})+\frac{\omega_{x}}{4}(\hat{x}^{2}+\hat{p}_{x}^{2})\nonumber \\
 & +\frac{\omega_{y}}{4}(\hat{y}^{2}+\hat{p}_{y}^{2})+g_{x}\hat{x}\hat{Z}_{L}+g_{y}\hat{y}\hat{Z}_{L},\label{eq:2DHamiltonian}
\end{alignat}
where $\omega_{x}$, $\omega_{y}$ are the frequencies of the two
harmonic motions, $\hat{x}$, $\hat{y}$ ($\hat{p}_{x}$, $\hat{p}_{y}$)
are the the position (momentum) observables, $g_{x}$, $g_{y}$ denote
the optomechanical couplings (see Eq.~\eqref{eq:couplings}), and
$\hat{Z}_{L}$ ($\hat{P}_{L}$) denote the amplitude (phase) quadrature
of the intracavity field.

The resulting equations of motion, including dissipation and Gaussian
noise baths acting on each mode, yield a set of linear coupled equations
which are represented in the well-known form: 
\begin{equation}
{\dot{{\bf X}}}={\bf A}{\bf X}+{\bf {\sqrt{\Gamma}}}{\bf X}_{\text{in}}(t)\label{Drift}
\end{equation}
where $\mathbf{A}$ is a drift matrix that includes dissipative terms
and $j$th element of the vector $(\mathbf{AX})^{(j)}=\frac{1}{i\hbar}[\mathbf{X}^{(j)},{\hat{H}}]-\frac{1}{2}(\mathbf{\Gamma X})^{(j)}$.
For our discussion we neglect the $z$-motion (but it is included
in quantum numerics), so can consider $\mathbf{A}$ as a $6\times6$
matrix. $\mathbf{X}$ is the vector of the mechanical and optical
modes, $\mathbf{X}=\begin{pmatrix}\hat{x},\,\hat{p}_{x},\,\hat{y},\,\hat{p}_{y},\,\hat{Z}_{L},\,\hat{P}_{L}\end{pmatrix}^{\textsf{T}}$,
$\sqrt{\Gamma}$ represents the diagonal matrix of damping coefficients
while ${\bf X}_{\text{in}}(t)$ represents the Gaussian noises (gas
collisions and optical shot noises).

To obtain the classical normal modes and frequencies of the system,
we calculated the eigenvalues and eigenvectors of ${\bf A}$ as a
function of the optical detuning $\Delta$ for the 2D $g_{x}=g_{y}$
case. It was sufficient for our classical analysis to consider the
case with dissipative terms set to zero (${\bf {\sqrt{\Gamma}}}=0$).

Such eigenmodes can be represented as unit vectors in the space spanned
by the tweezer modes $\hat{x}$, $\hat{\ensuremath{y}}$, and $\hat{Z}_{L}$,
i.e., as spherical polar angles on a Bloch sphere (see Fig.~\eqref{spins}
(a)). As the detuning is varied from $\Delta=-\infty$ to $\Delta\sim0$
we represented the resulting trajectory traced by each eigenmode using
the corresponding spherical polar angles -- Fig.~\ref{spins}(b)
plots the trajectories as a function of detuning $\Delta$ for two
realistic scenarios. In the first case (orange lines) we employ the
experimental parameters of ~\citep{delic2020cooling} and thus a
more elliptical tweezer trap $w_{x}=0.6\mu$m, $w_{y}=0.705\mu$m
where tweezer frequencies differ by about $16\%$. In the second case
(red lines) we set $w_{x}=0.68\mu$m, so now the frequencies are near-degenerate
(near circular trap) and differ by about $3.5\%$.

The plots in Fig.\ref{spins}(b) show that for $\Delta=-\infty$,
the modes tend to the uncoupled $\hat{x}$, $\hat{\ensuremath{y}}$,
$\hat{Z}_{L}$ modes (green arrows). The red lines (circular trap)
approach closely the `dark/bright' modes of Eq.\eqref{eq:dark}, which
represents the $g_{x}=g_{y}$, $\omega_{x}=\omega_{y}$ limit. We
see the dark mode simply rotates on the ``equator'' ($\theta=\pi/2$),
from $\phi=0$ at $\Delta=-\infty$ through to $\phi=-\pi/4$ at the
centre of the crossing $\-\Delta=\omega_{x}\simeq\omega_{y}$. For
the elliptical trap of the experiments ~\citep{delic2020cooling}
however, the `dark' mode (top panel) still rotates to $\phi=-\pi/4$
at the centre of the crossing but mixes appreciably with the optical
mode and $\theta\simeq0.6\pi$.

\begin{figure}[t]
\includegraphics[width=0.8\columnwidth]{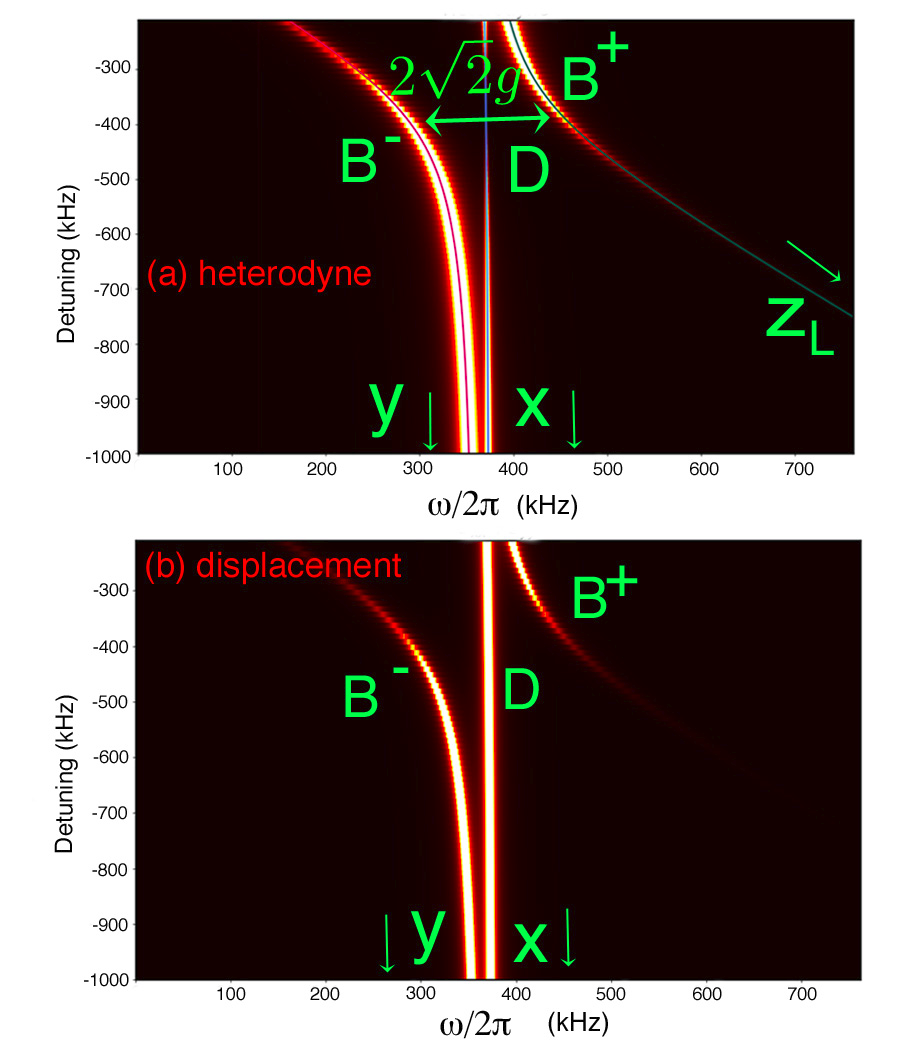} \caption{\textbf{(a)} Illustrates heterodyne PSD $S_{\text{het}}(\omega)$,
in regimes where the $\hat{x},\hat{y}$ and optical modes hybridise
to form bright ($\text{\ensuremath{\hat{B}}}^{\pm}$) and dark ($\hat{D}$)
modes, i.e., where $g\gg|\omega_{x}-\omega_{y}|,\kappa,\gamma$. The
solid lines overlaid are the classical modes. The central dark mode
reaches near zero amplitude at the centre of the crossing. The bright
modes show an avoided crossing of width enhanced by a factor of $\sqrt{2}$.
\textbf{(b)} In contrast, the displacement PSD, $S_{xx}(\omega)+S_{yy}(\omega)$
show that both dark and bright modes are uncooled and remain hot.
\label{Fig2:BDmodes}}
\end{figure}

The more general case of $g_{x}\neq g_{y}$ (and arbitrary detuning)
does not eliminate the bright-dark mode structure. In contrast lifting
the $\omega_{x}\sim\omega_{y}$ degeneracy has a pronounced effect
-- the bright/dark modes mix and very different trajectories are
obtained. Ultimately this would lead to a decoupling to two independent
level crossings, with the associated disadvantage that both modes
might no longer be resonant simultaneously -- this case depends on
$\kappa$ and is investigated below.

Fig.\ref{Fig2:BDmodes}(a) illustrates the characteristics of a heterodyne
PSD $S_{\text{het}}(\omega)$, in regimes of bright dark-modes, where
$g\gg|\omega_{x}-\omega_{y}|,\kappa,\gamma$ so dissipation is very
low. Specifically, the parameters are set close to experiment of ~\citep{delic2020cooling}
but with $\theta_{tw}=\pi/4$, $\kappa\to\kappa/10$ and $|\omega_{x}-\omega_{y}|\to|\omega_{x}-\omega_{y}|/4$.
The figure illustrates the typical structure of a 3-level crossing
with a coupling in the form of Eq.~\eqref{eq:Sxmatrix} ( two degenerate
modes $\hat{x},\hat{y}$ couple indirectly via a third). It illustrates
however also the difficulty of the usual procedure for thermometry
in optomechanics. Both modes are hot, and moreover the usual normalisation
used to relate the heterodyne measured PSD to the underlying displacement
spectra gives very poor results.

\begin{table}
\begin{tabular}{|c|c|c|}
\hline 
parameter  & symbol  & value\tabularnewline
\hline 
\hline 
gas pressure  & $p$  & $10^{-6}\text{mbar}$\tabularnewline
\hline 
gas temperature  & $T$  & $300\,\text{K}$\tabularnewline
\hline 
cavity decay rate  & $\kappa/2\pi$  & $193\,\text{kHz}$\tabularnewline
\hline 
cavity length  & $L$  & $1.07\text{cm}$\tabularnewline
\hline 
cavity waist  & $w_{c}$  & $41.1\,\text{\ensuremath{\mu}m}$\tabularnewline
\hline 
optical wavelength  & $\lambda$  & $1064\,\text{nm}$\tabularnewline
\hline 
silica density  & $\rho$  & $2000\,\text{kg}\text{m}^{-3}$\tabularnewline
\hline 
particle radius  & $R$  & $71.5\,\text{nm}$\tabularnewline
\hline 
input tweezer power  & $P_{\text{tw}}$  & $400\,\text{mW}$\tabularnewline
\hline 
x tweezer waist  & $w_{x}$  & $0.600\,\text{\ensuremath{\mu}m}$\tabularnewline
\hline 
y tweezer waist  & $w_{y}$  & $0.705\,\text{\ensuremath{\mu}m}$\tabularnewline
\hline 
\end{tabular}\caption{Values used in numerical simulations adapted from the experiment~\cite{delic2020cooling}.
The last four parameters ($R$, $P_{\text{tw}}$, $w_{x}$, $w_{y}$)
are the nominal values which are modified in the numerical simulations
to scan the different regimes of \emph{three-way }hybridization and
2D cooling/thermometry (see Figs.~1-6).\label{tab:Values-used-in}}
\end{table}

\section{2D motion modeling\label{sec:analysis} }

In \citep{torovs2020quantum} the Langevin equations for the full
3D problem was solved, both for the full non-linear trapping potentials
as well as considering linearisations about equilibrium values. For
the linearised case general expressions of Quantum Linear theory (QLT)
for the optical modes were obtained as well as the mechanical spectra
$\hat{x}^{\text{3D}},\hat{y}^{\text{3D}}$ and $\hat{z}^{\text{3D}}$,
including hybridisation and encompassing quantum regimes.

From these, power spectral densities (PSDs) for $S_{xx}(\omega)=\langle|\hat{x}^{\text{3D}}(\omega)|^{2}\rangle$
or $S_{yy}(\omega)=\langle|\hat{y}^{\text{3D}}(\omega)|^{2}\rangle$
were calculated and thence phonon occupancies are related to the area
under the PSD curve~\citep{bowen2015quantum}: 
\begin{equation}
n_{j}=\frac{1}{2\pi}\int_{-\infty}^{\infty}S_{jj}(\omega)d\omega-\frac{1}{2}\label{Phon}
\end{equation}
for $j=x,y,z$ and compared with optical (heterodyne) PSDs in order
to understand the experimental measurements.

However, in \citep{torovs2020quantum}, the analysis of quantum cooling
for particles trapped at the node of the cavity, focussed on quasi-1D
experimental scenarios of ~\citep{delic2020cooling} where only a
single mode was strong-coupled to the light. Hybridisation with weak-coupled
modes, leading to sympathetic cooling or heating was investigated.
In the present work we go beyond \citep{torovs2020quantum} to investigate
the case of two strong-coupled modes, where non-trivial 2D physics
arises, including the formation of dark modes.

In addition, we consider also scattering effects not included in \citep{torovs2020quantum}.
The Hamiltonian of the system, Eq.~\eqref{eq:2DHamiltonian} is a
special case of the Hamiltonians discussed in~\cite{delic2019cavity,torovs2020quantum}
where the nanoparticle equilibrium position was primarily determined
by the gradient force ($\sim70\text{nm}$ particles). Here we however
consider also substantially larger particles ($\sim100\text{nm}$)
where the scattering force must be taken into account. In particular,
the latter displaces the nanoparticle equilibrium position which leads
to new couplings of $\hat{x}$ and $\hat{y}$ to the phase quadrature
of light, $\hat{P}$ (in addition, to the coupling to the amplitude
quadrature, $\hat{Z}$). In Appendix \ref{sec:Notes-on-scattering}
we show that the Hamiltonian can be transformed back to the form in
Eq.~\eqref{eq:2DHamiltonian} by introducing the rotated optical
quadratures $\hat{Z}_{L}$ ($\hat{P}_{L}$) with the rotation angle
depending on the size of the nanoparticle. In short, all of the results
from \cite{delic2019cavity,torovs2020quantum} remain valid even when
the scattering force is non-negligible (but still in the Rayleigh
regime) as long as we formally replace $\hat{Z}$,$\hat{P}$ with
the rotated optical quadratures $\hat{Z}_{L}$,$\hat{P}_{L}$.

For our 2D analysis, we assume the particle is trapped at the cavity
node, i.e. $\phi_{\text{tw}}=\pi/2$ that minimises deleterious optical
heating, and ask: what is the optimal angle $\theta_{\text{tw}}$
between the tweezer polarization axis and the cavity symmetry axis
for efficient 2D cooling? The latter controls the couplings $g_{x}$,
$g_{y}$, and the most natural choice is given by $\theta_{\text{tw}}\sim\pi/4$
where $g_{x}\sim g_{y}$ -- this maximizes the cooperativities of
both the $x$ and $y$ motions. In addition, we have the freedom in
choosing the detuning, $\Delta$ ($\Delta<0$ is red-detuned) --
in first instance this can be set to $-\Delta=(\omega_{x}+\omega_{y})/2$.
In particular, the perfectly degenerate case, $\omega_{x}=\omega_{y}$,
where we have the exact relation $g_{x}=g_{y}$ seems the most natural
configuration for 2D cooling -- however, we will show this is not
the case, and non-degenerate frequencies are necessary for efficient
2D cooling.

The 2D equations of motion ${\dot{{\bf X}}}={\bf A}{\bf X}$ in Eq.~\eqref{Drift},
explicitly, are given by: 
\begin{alignat}{2}
\dot{\hat{x}} & =\omega_{x}\hat{p}_{x}, & \dot{\hat{p}}_{x} & =-\omega_{x}\hat{x}-2g_{x}\hat{Z}_{L},\label{eq:xdot}\\
\dot{\hat{y}} & =\omega_{y}\hat{p}_{y}, & \dot{\hat{p}}_{y} & =-\omega_{y}\hat{x}_{d}-2g_{y}\hat{Z}_{L},\\
\dot{\hat{Z}}_{L} & =-\Delta\hat{P}_{L},\qquad & \dot{\hat{P}}_{L} & =\Delta\hat{Z}_{L}-2g_{x}\hat{x}-2g_{y}\hat{y},\label{eq:PLdot}
\end{alignat}
where we have for simplicity of presentation omitted the non-conservative
terms (damping terms and input noises). We have the optical quadratures,
$\hat{Z}_{L}=\hat{a}+\hat{a}^{\dagger}$ and $\hat{P}_{L}=i(\hat{a}^{\dagger}-\hat{a})$,
x mechanical quadratures, $\hat{x}=\hat{b}_{x}+\hat{b}_{x}^{\dagger}$
and $\hat{p}_{x}=i(\hat{b}_{x}^{\dagger}-\hat{b}_{x})$, and the y
mechanical quadratures, $\hat{y}=\hat{b}_{y}+\hat{b}_{y}^{\dagger}$
and $\hat{p}_{y}=i(\hat{b}_{y}^{\dagger}-\hat{b}_{y})$. We can express
the equations for the modes of the 2D problem in Fourier space: 
\begin{alignat}{1}
\hat{x}(\omega)= & J_{xZ}(\omega)\hat{Z}_{L}(\omega)+\tilde{x}_{\text{in}}(\omega),\label{eq:xF}\\
\hat{y}(\omega)= & J_{yZ}(\omega)\hat{Z}_{L}(\omega)+\tilde{y}_{\text{in}}(\omega),\label{eq:yF}\\
\hat{Z}_{L}(\omega)= & J_{Zx}(\omega)\hat{x}(\omega)+J_{Zy}(\omega)\hat{y}(\omega)+\tilde{Z}_{L,\text{in}}(\omega),\label{eq:ZLF}
\end{alignat}
which can be solved in closed form as shown in \citep{torovs2020quantum}.
The solutions will be labelled as $\hat{x}^{\text{2D}}(\omega)$,
$\hat{y}^{\text{2D}}(\omega)$ and $\hat{Z}_{L}(\omega)$ (which in general depend on all three input noises $\tilde{x}_{\text{in}}(\omega)$,
$\tilde{y}_{\text{in}}(\omega)$ and $\tilde{Z}_{L,\text{in}}(\omega)$).
We recall that our calculations fully include the $z$ mechanical
motion. However the 2D analysis below is closely matched by the full
3D theory and thus we can make the identifications, $\hat{x}^{\text{2D}}(\omega)\equiv\hat{x}^{\text{3D}}(\omega)$
and $\hat{y}^{\text{2D}}(\omega)\equiv\hat{y}^{\text{2D}}(\omega)$,
for the hybridised modes of the system. The frequency dependent coupling
coefficients are given by

\begin{alignat}{1}
J_{jZ}(\omega) & =-2g_{j}\chi_{j}(\omega),\label{eq:jZ}\\
J_{Zj}(\omega) & =-ig_{b}\eta(\omega),\label{eq:Zj}
\end{alignat}
where $j=x,y$, the susceptibilities are given by

\begin{alignat}{1}
\chi_{j}(\omega) & =\frac{\omega_{j}}{-\omega^{2}+\omega_{j}^{2}-i\omega_{j}\gamma},\label{eq:mechanical}\\
\eta(\omega) & =\frac{1}{-i(\omega+\Delta)+\frac{\kappa}{2}}-\frac{1}{i(-\omega+\Delta)+\frac{\kappa}{2}},\label{eq:optical}
\end{alignat}
and $\kappa$ ($\gamma$) is the cavity decay rate (gas damping).

Eqs.~\eqref{eq:xF}-\eqref{eq:ZLF} form a system of coupled equations:
the solutions $\hat{x}(\omega)$, $\hat{y}(\omega)$, $\hat{a}(\omega)$
are function of the input noises $\tilde{x}_{\text{in}}(\omega)$,
$\tilde{y}_{\text{in}}(\omega)$, $\tilde{a}_{\text{in}}(\omega)$.
In addition to gas collisions and photon shot-noise we also include
recoil heating in the model by adding additional terms to $\tilde{x}_{\text{in}}(\omega)$
and $\tilde{y}_{\text{in}}(\omega)$~\citep{jain2016direct,delic2019cavity,seberson2020distribution}.
The latter can become relevant even at pressure $p\sim10^{-6}\text{mbar}$
when we scan over large values of the couplings $g_{x}$,$g_{y}\apprge100\text{kHz}$
as we increase the size of the nanoparticle/laser power.

\section{2D cooling -- numerical results\label{sec:results}}

We define the mechanical 2D phonon occupancy as

\begin{equation}
\hat{n}^{\text{(2D)}}=\hat{n}_{x}+\hat{n}_{y}\,,\label{eq:n2d-2}
\end{equation}
where $\hat{n}_{x}=\hat{b}_{x}^{\dagger}\hat{b}_{x}$ and $\hat{n}_{y}=\hat{b}_{y}^{\dagger}\hat{b}_{y}$.
The latter were calculated using Quantum Linear Theory (QLT) by numerically
integrating the associated power spectral densities (PSDs), $S_{xx}$
and $S_{yy}$ (see Sec.~\ref{sec:analysis}) and comparing with phonon
occupancies inferred from optical (heterodyne) spectra. We emphasise
that the numerics are all full 3D, but our analysis -- for physical
insight -- considers the reduced 2D motion.

In the following we consider the condition $\hat{n}^{\text{(2D)}}<1$
as a threshold value for 2D motional ground state cooling. Alternatively
we could have required the less restrictive conditions $\hat{n}_{x},\hat{n}_{y}<1$,
i.e. the two motions are separately in the ground state.

\subsection{Optimal frequency difference\label{subsec:Optimal-frequency-difference}}

\begin{figure}[t]
\includegraphics[width=1\columnwidth]{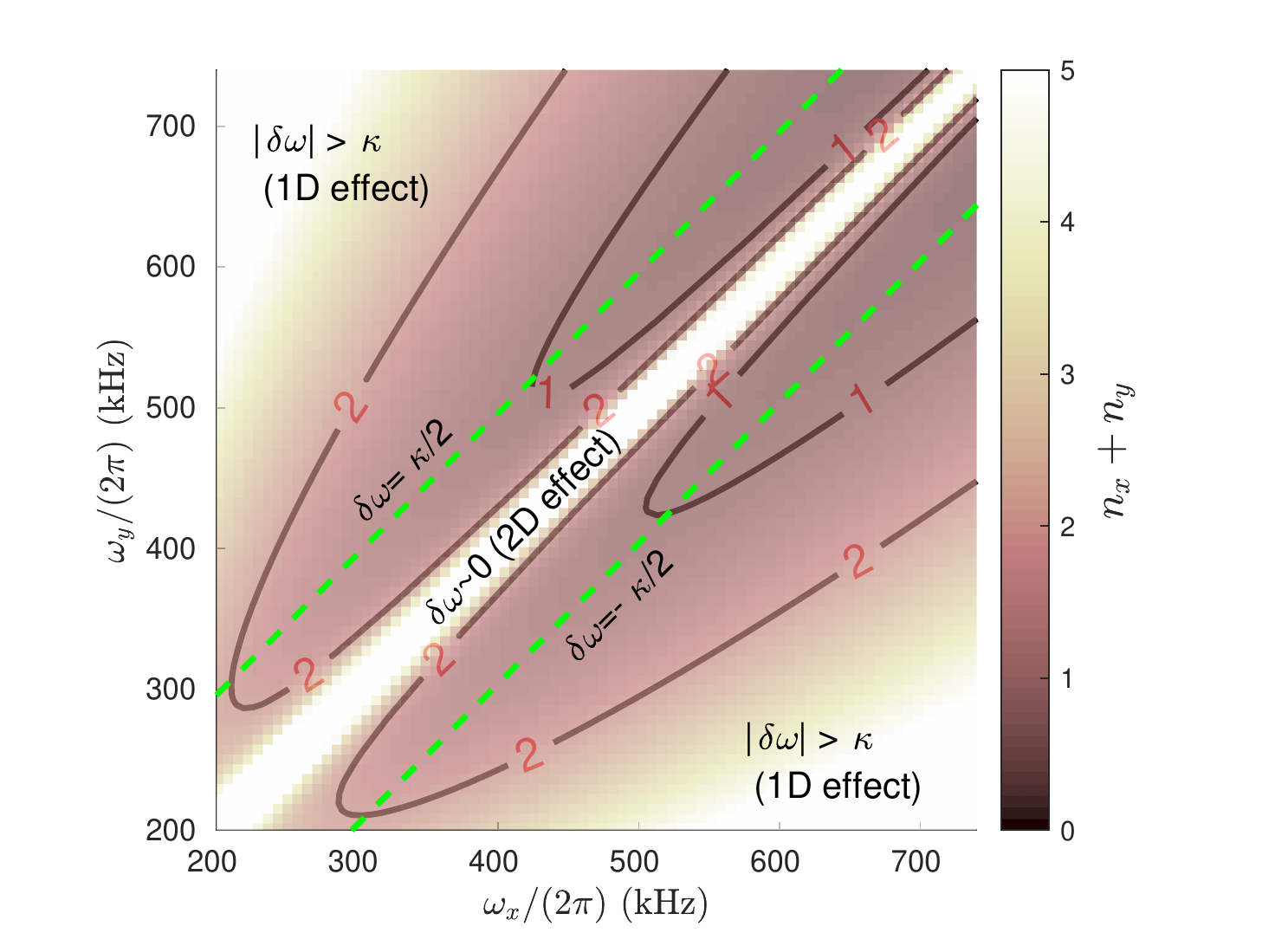}

\caption{Numerical simulation of 2D phonon occupancy as a function of the tweezer
frequencies $\omega_{x}$ and $\omega_{y}$. The values are set as
in \cite{delic2020cooling} but $\theta=\pi/4$ and with variable
tweezer waists along the x, y-axis. Cooling becomes ineffective in
two regimes of $\delta\omega=\vert\omega_{x}-\omega_{y}|$: (i) when
the trap is near circular and a decoupled dark mode is formed (diagonal
white strip) or (ii) the frequency difference $\delta\omega$ is too
large compared to the cavity decay rate $\kappa$ (white regions in
top left and bottom right corners). Cooling is optimal for intermediate
frequency differences when $\delta\omega$ is close to $\kappa/2$.\label{fig:phonons_frequencies}}
\end{figure}

The first question we address is what is the optimal frequency difference,
$\delta\omega=\vert\omega_{x}-\omega_{y}\vert$, in order to achieve
the lowest combined phonon occupancy $\hat{n}^{\text{(2D)}}$. For
concreteness we will consider the parameters from \cite{delic2020cooling}
(but now with $\theta=\pi/4$) and vary the two waists of the tweezer
beam, $w_{x}$ and $w_{y}$, to scan over the frequencies, $\omega_{x}$,
$\omega_{y}$. We find that when $\vert\omega_{x}-\omega_{y}\vert>\kappa$
simultaneous cooling of the $\hat{x}$ and $\hat{y}$ modes becomes
ineffective -- we either cool $x$-motion or $y$-motion, but cannot
cool both effectively. More surprisingly, we find that when $\vert\omega_{x}-\omega_{y}\vert\sim0$
cooling becomes again ineffective. The optimal frequency difference
for efficient 2D cooling is near the midpoint value -- when $\vert\omega_{x}-\omega_{y}\vert\sim\frac{\kappa}{2}$
with the detuning set to $-\Delta\sim(\omega_{x}+\omega_{y})/2$ (see
Fig.~\ref{fig:phonons_frequencies}).

We can understand qualitatively the reason for the optimal frequency
difference $\vert\omega_{x}-\omega_{y}\vert\sim\frac{\kappa}{2}$
by calculating 2D optomechanical cooling formula:

\begin{equation}
\Gamma_{\text{opt,j}}\equiv\text{Im}\left[\frac{2ig_{j}^{2}\eta(\omega_{j})}{1-2ig_{k}^{2}\chi_{k}(\omega_{j})\eta(\omega_{j})}\right],\label{eq:gammaoptj}
\end{equation}
where $j=x,k=y$ or $j=y,k=x$ (from Eqs.~\eqref{eq:xF}-\eqref{eq:jZ}
one readily finds the optomechanical cooling formula by calculating
the imaginary part of the self-energy~\citep{marquardt2007quantum,torovs2020quantum}).
Let us consider some special cases. Suppose first that $g_{k}\sim0$
such that we have $\Gamma_{\text{opt,j}}\sim\text{Im}\left[2ig_{j}^{2}\eta(\omega_{j})\right]$
-- the latter is the usual optomechanical cooling rate which further
reduces to $\Gamma_{\text{opt,j}}\sim4g_{j}^{2}/\kappa$. The numerator
can be thus associated with the cooling rate from standard 1D cavity
optomechanics. On the other hand the denominator depends only on the
coupling to the other degree of freedom, $\sim g_{k}$, and is thus
a genuinely 2D effect affecting the j-motion.

We are primarily interested in the configuration where both $\Gamma_{\text{opt,x}}$
and $\Gamma_{\text{opt,y}}$ are large. Let us start by considering
the perfectly degenerate case, $\omega_{x}=\omega_{y}=-\Delta$ with
$g=g_{x}\sim g_{y}$. Assuming the regime of strong cooperativity
we find that the optomechanical rate reduces to the simple expression
$\Gamma_{\text{opt,j}}\sim\gamma$ . The gas damping, $\gamma$, is
tiny at the relevant pressures, and thus we are left only with a negligible
optomechanical cooling rate -- here $\gamma$ arises from the denominator
in Eq.~(\ref{eq:gammaoptj}), i.e. from the mechanical susceptibility
$\chi_{k}(\omega_{j})$ defined in Eq.~\eqref{eq:mechanical}, and
thus the strong suppression of the optomechanical cooling rate can
be identified as a 2D effect. Loosely speaking, the energy that is
extracted from the $x$-motion ($y$-motion) is immediately fed back
to the $y$-motion ($x$-motion) with the optical field mediating
this transition. In order to achieve any 2D cooling we thus require
some degree of asymmetry, $\omega_{x}\neq\omega_{y}$, in order to
disrupt the near-perfect exchange of energy between $\hat{x}$ and
$\hat{y}$ via the optical field, and allow the latter to instead
carry the energy away from the system.

We finally note that lowering the finesse does not necessarily improve
2D cooling. This is captured by the optomechanical cooling formula
in Eq.~\eqref{eq:gammaoptj} through the optical susceptibility $\eta$
defined in Eq.~\eqref{eq:optical}: on the one hand, when we decrease
the value of $\kappa$ we enhance the 1D cooling channel (numerator),
but, on the other hand, we also amplify the 2D heating channel (denominator).

\subsection{Optimal particle size\label{subsec:Optimal-particle-size}}

\begin{figure}[t]
\includegraphics[width=1\columnwidth]{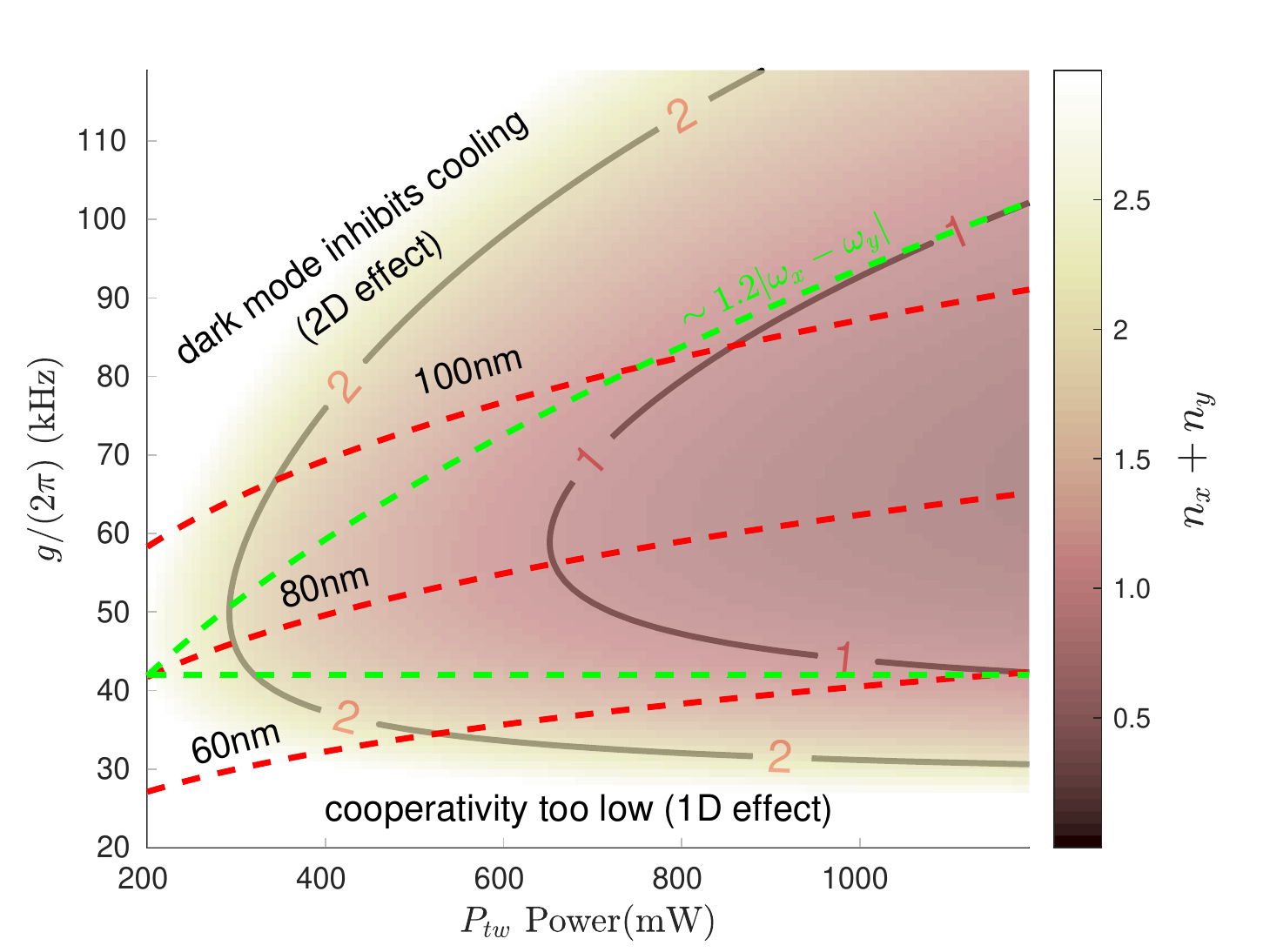}

\caption{Numerical simulation of 2D phonon occupancy as a function of input
power ($P_{\text{tw}}$) and mean optomechanical coupling ($g\equiv(g_{x}+g_{y})/2$).
The values are set as in \cite{delic2020cooling} but $\theta=\pi/4$
and variable power and particle radius (which sets the optomechanical
couplings). Cooling becomes ineffective if the cooperativity is too
low (below the lower green dashed line) as well as if decoupled dark
modes are formed (above the upper green dashed line). The red dashed
lines plot $g$ for different values of the particle radius, $R$.
We can thus extract the optimal particle size that allows efficient
2D cooling at moderate powers -- cooling to the 2D motional ground
state, $n_{x}+n_{y}<1$, is feasible already for a $\sim80\ \text{nm}$
particle at $P_{\text{tw}}\sim700\ \text{mW}$. \label{fig:phonons_particle}}
\end{figure}

For concreteness we will consider the parameters from \cite{delic2020cooling}
(but now with $\theta=\pi/4$) which is close to the optimal regime
$\vert\omega_{x}-\omega_{y}\vert\apprle\frac{\kappa}{2}$ (see Sec.~\ref{subsec:Optimal-frequency-difference}),
with the detuning set to $-\Delta\sim(\omega_{x}+\omega_{y})/2$ .

For a given experimental implementation the tweezer power, $P_{tw}$,
is a parameter that can be varied readily. However to minimise deleterious
optical heating effects, it is preferrable to use the lowest power
that meets experimental requirements. We restrict ourselves to $P_{tw}\lesssim1$W.
Nanoparticles of different radii $R$ may also be selected.

We express the relevant parameters for 2D cooling, $\omega_{x}$,
$\omega_{y}$, $g_{x}$, $g_{y}$ as a function of $P_{tw}$ and $R$.
We then find that if the particle size is too small ($\apprle60\text{nm}$)
the cooperativity remains low and one is limited to values above $n_{x}+n_{y}\sim1$
-- this is analogous to the requirement for 1D ground state cooling.
However, if the particle size is too large ($\apprge100\text{nm}$)
then cooling becomes again ineffective when $g_{x}$, $g_{y}\apprge\vert\omega_{x}-\omega_{y}\vert$.
We find that there is a ``Goldilocks zone'' with the optimal particle
size $\sim80\ \text{nm}$ (see Fig.~\ref{fig:phonons_particle}).

We can understand qualitatively the reason for the optimal particle
size by looking again at the 2D optomechanical cooling formula in
Eq.~\eqref{eq:gammaoptj}. We first note that $g_{j}\propto R^{3/2}$
and that $g_{j}\propto P_{tw}^{1/4}$ hence $g_{j}\propto R^{3/2}P_{tw}^{1/4}$.
For small (large) values of $g_{j}$ the numerator (denominator) in
Eq.~(\ref{eq:gammaoptj}) is small (large) and cooling becomes inhibited
-- this illustrates how the ``Goldilocks zone'' emerges from the
competition of the 1D effect in the numerator with the 2D effect in
the denominator. In particular, the condition $g_{j}\apprle\vert\omega_{x}-\omega_{y}\vert$
emerges from the denominator in Eq.~(\ref{eq:gammaoptj}) -- we
have $\chi_{k}(\omega_{j})\sim\vert\omega_{x}-\omega_{y}\vert^{-1}$
as well as $\eta(\omega_{j})\sim\vert\omega_{x}-\omega_{y}\vert^{-1}$
(since $-\Delta\sim(\omega_{x}+\omega_{y})/2$) -- and hence the
the denominator remains suppressed if $\vert g_{k}\vert\apprle\vert\omega_{x}-\omega_{y}\vert$,
i.e. cooling is not inhibited by the 2D hybridisation effect.

\subsection{Reliable thermometry}

\begin{figure}
\centering{}\includegraphics[width=1\columnwidth]{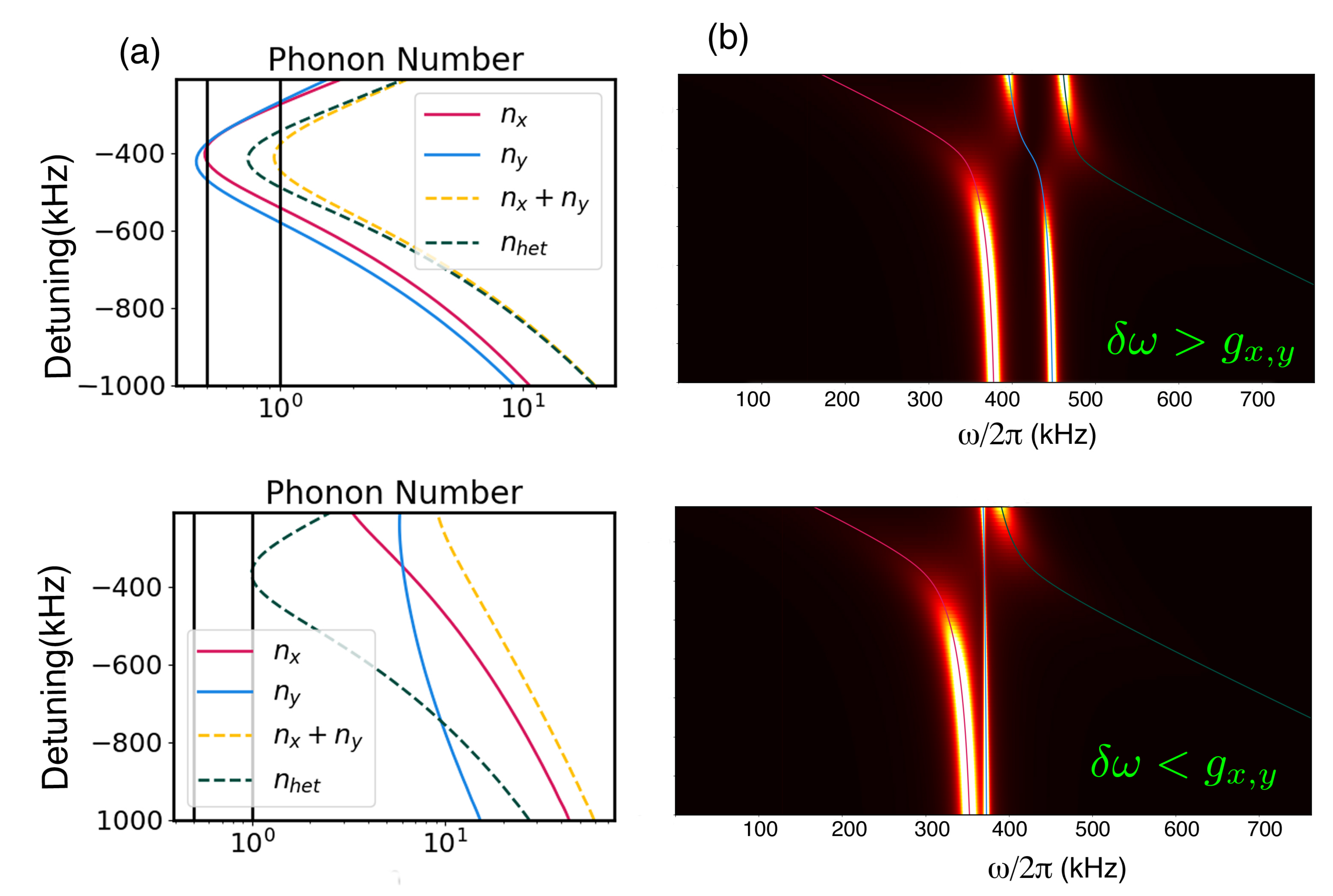} \caption{\textbf{(a)} Compares phonon occupancies for an elliptical trap used
in the quantum cooling experiments \cite{delic2020cooling}, $\delta\omega=|\omega_{x}-\omega_{y}|>g_{x}\sim g_{y}$
(upper) with occupancies for a near-circular trap with the same parameters
but $\delta\omega$ smaller so that $|\omega_{x}-\omega_{y}|<g_{x}\sim g_{y}$
(lower). The elliptical trap allows for 2D ground state cooling and
the rescaled heterodyne follows occupancies closely, facilitating
thermometry. For the near-circular trap, the modes remain hot and
it is difficult to extract occupancies from the optical detection
by the usual methods. The particle is positioned at a node (intensity
minimum), $\theta_{\text{tw}}=\pi/4$, $R=80$ nm, input power $P_{in}=0.8$
W, and $\kappa=193$ kHz. \textbf{(b)} Corresponding heterodyne PSDs,
with the classical modes overlaid. For the elliptical trap in the
upper panels when the detuning is set to $-\Delta=400$ kHz the modes
are cooled to $n_{x}+n_{y}<1$. \label{fig:phononsAndSpectra}}
\end{figure}

In the previous sections we have shown that there exists an optimal
experimental configuration (Sec.~\ref{subsec:Optimal-frequency-difference})
and particle size (Sec.~\ref{subsec:Optimal-particle-size}) to achieve
simultaneous cooling of both $x$ any $y$ motions. However, inferring
phonon occupancies from optically detected spectra in the presence
of hybridisation is not straightforward. Here we show that the same
experimental configuration that allows for optimal 2D cooling also
allows for reliable thermometry.

Experiments exploiting heterodyne detection have access only to the
optical field, $\hat{a}=\frac{1}{2}(\hat{Z}_{L}+i\hat{P}_{L})$, from
which one then extracts the the mechanical displacement spectra. In
particular, the heterodyne PSD is given by~\citep{bowen2015quantum}
\begin{equation}
S_{\text{het}}(\omega)\equiv S_{a_{\text{out}}a_{\text{out}}^{\dagger}}(\Delta_{\text{LO}}+\omega)+S_{a_{\text{out}}^{\dagger}a_{\text{out}}}(\Delta_{\text{LO}}-\omega),\label{eq:het}
\end{equation}
where $\Delta_{\text{LO}}$ is the detuning of the local oscillator,
and $\hat{a}_{\text{out}}=\hat{a}_{\text{in}}-\sqrt{\kappa}\hat{a}$
is the output field.

In the presence of hybridisation and spectral overlaps, extracting
displacement spectra $S_{xx}(\omega)$ and $S_{yy}(\omega)$, from
the experimental heterodyne spectra, $S_{\text{het}}(\omega)$, becomes
less straightforward. However, in the optimal case for 2D cooling
(see Sec.~\ref{subsec:Optimal-frequency-difference}) -- when $g_{x}\simeq g_{y}\simeq g$
, we can write: 
\begin{equation}
\hat{a}(\omega)\simeq\eta(\omega)g[\hat{x}^{\text{2D}}(\omega)+\hat{y}^{\text{2D}}(\omega)].\label{eq:afield}
\end{equation}

The corresponding heterodyne PSD (from Eq.\ref{eq:het}) will have
contributions not only from independent $x,y$ contributions, but
also from interferences. Thermometry is greatly simplified when we
neglect interference effects and are able to write:

\begin{equation}
\frac{S_{\text{het}}(\omega)}{|\eta(\omega)|^{2}g^{2}}\simeq S_{xx}(\omega)+S_{yy}(\omega),\label{eq:formula}
\end{equation}
which can be seen as the 2D extension of the familiar textbook relation
arising in the 1D case. Clearly more elliptical traps, with less spectral
overlap between the $x,y$ terms would be expected to minimise interference
contributions.

We test the approximation in Eq.~\eqref{eq:formula} in Fig.~\ref{fig:phononsAndSpectra}:
in panels (a) we compare the extracted phonon occupancies (using the
heterodyne spectra) with the actual ones, and in panels (b) we show
the PSDs for the heterodyne signal with the classical modes overlaid.
The upper panels show the result for an elliptical trap lie within
the ``Goldilocks zone'' where frequencies are sufficiently far apart
so that both modes are in the quantum regime and moreover, the phonon
occupancies inferred from rescaled heterodyne PSD agree reasonably
well with those obtained from integrating $S_{xx}(\omega)+S_{yy}(\omega)$,
in contrast with the near-circular trap, which lies outside this zone.

In the latter case (near-circular trap) the modes remain hot and more
complicated methods would be required to infer mode occupancies from
the optically detected signal.

\subsection{Understanding 2D cooling in terms of geometric bright/dark modes\label{subsec:Understanding-2D-cooling}}

In our effective 2D system, $\hat{x}^{\text{2D}}$ and $\hat{y}^{\text{2D}}$
are the mechanical modes. As shown in the previous section, they in
general arise from hybridisation of $\hat{x}$, $\hat{y}$ modes and,
as we consider regimes of strong-coupling, they also involve hybridisation
with the optical mode $\hat{Z}_{L}$. Unhybridised $\hat{x}$, $\hat{y}$
mechanical modes would correspond physically to motions along the
tweezer $x,y$ axes, respectively (such as in the case the coupling
to the cavity mode would be vanishingly small and we would thus have
two completely decoupled mechanical motions).

However, there is another pair of mechanical modes that naturally
arises in the coherent scattering setup: these are the bright (dark)
modes. We show below that, in our system, the former have an interesting
and useful geometric interpretation in terms of modes corresponding
to the motion along (orthogonal) to the cavity axis; together with
the cavity mode, $\hat{Z}_{L}$, they form the \emph{cavity-based/geometric
}modes.

The transformation from the tweezer-based modes to the geometric/cavity-based
modes might, in first instance, be understood as a pure 2D rotation
in the $x$-$y$ plane by applying a rotation of angle $\theta_{\text{tw}}$.
However, care is required when we consider traps with significant
ellipticity, where $\omega_{x}\neq\omega_{y}$, where the zero-point
motions distort the 2D rotation as we will now show. We start by considering
the rotated reference frame where the first (second) axis of the reference
frame is parallel (orthogonal) to the cavity-axis (see Fig.~\ref{Fig1}).
Specifically, we transform to such ``cavity reference frame'' by
applying a rotation of angle $\theta_{\text{tw}}$:

\begin{equation}
\left[\begin{array}{c}
X^{\text{(c)}}\\
Y^{\text{(c)}}
\end{array}\right]=\underbrace{\left[\begin{array}{cc}
\text{sin}\theta_{\text{tw}} & \text{cos}\theta_{\text{tw}}\\
-\text{cos}\theta_{\text{tw}} & \text{sin}\theta_{\text{tw}}
\end{array}\right]}_{\mathbb{\equiv\mathcal{R}}(\theta_{\text{tw}})}\left[\begin{array}{c}
X\\
Y
\end{array}\right],\label{eq:rotation}
\end{equation}
where $X,Y$ ($X^{\text{(c)}},Y^{\text{(c)}}$) are the coordinates
in tweezer (cavity) reference frame. The rotation in Eq.~(\ref{eq:rotation})
in term induces a transformation of the canonical (adimensional) modes
$\hat{x}$ and $\hat{y}$. In a nutshell, one first rotates the corresponding
physical positions ($\hat{X}$ and $\hat{Y}$) to obtain the transformed
physical positions ($\hat{X}_{b}$ and $\hat{X}_{d}$) and thence
defines the transformed canonical positions ($\hat{x}_{b}$ and $\hat{x}_{d}$)
by rescaling them with the transformed zero-point motions. Specifically,
we perform the following transformations in consecutive order:

\begin{alignat}{1}
\left[\begin{array}{c}
\hat{x}\\
\hat{y}
\end{array}\right] & \rightarrow\left[\begin{array}{c}
\hat{X}\\
\hat{Y}
\end{array}\right]\equiv\left[\begin{array}{c}
x_{\text{zpf }}\hat{x}\\
y_{\text{zpf}}\hat{y}
\end{array}\right],\label{eq:first}\\
\left[\begin{array}{c}
\hat{X}\\
\hat{Y}
\end{array}\right] & \rightarrow\left[\begin{array}{c}
\hat{X}_{b}\\
\hat{X}_{d}
\end{array}\right]\equiv\mathbb{\mathcal{R}}(\theta_{\text{tw}})\left[\begin{array}{c}
\hat{X}\\
\hat{Y}
\end{array}\right],\\
\left[\begin{array}{c}
\hat{X}_{b}\\
\hat{X}_{d}
\end{array}\right] & \rightarrow\left[\begin{array}{c}
\hat{x}_{b}\\
\hat{x}_{d}
\end{array}\right]\equiv\left[\begin{array}{c}
\hat{X}_{b}/x_{b,\text{zpf }}\\
\hat{X}_{d}/x_{d,\text{zpf }}
\end{array}\right],\label{eq:second}
\end{alignat}
where $x_{j,\text{zpf }}=\sqrt{\frac{\hbar}{2m\omega_{j}}}$ ($\omega_{j}$)
is the zero-point motion (frequency) along the $j=b,d$ axis (for
the full details of the derivation see Appendix~\ref{sec:2dRotations}
and \ref{sec:brightdark}).

The Hamiltonian from Eq.~\eqref{eq:2dOpto} in terms of the new rotated
coordinates reduces to 
\begin{equation}
\frac{\hat{V}_{\text{int}}}{\hbar}=g_{bd}\hat{x}_{b}\hat{x}_{d}+g_{b}\hat{x}_{b}\hat{Z}_{L},\label{eq:HamiltonianBD}
\end{equation}
and thus only the mode $\hat{x}_{b}$ is coupled to the light field
while $\hat{x}_{d}$ is completely decoupled -- we will refer to
them as the \emph{geometric} \emph{bright} and \emph{dark} mode, respectively.
A similar coupling to the above was obtained in a previous experimental
study of hybridisation between two mechanical modes~\cite{shkarin2014optically}
leading to bright/dark modes (albeit not in a strong coupling regime
and without strong cooling).

However, what is new is that in the present case the resulting modes,
$\hat{x}_{b}$ and $\hat{x}_{d}$, have a simple geometric interpretation
as the motion along and orthogonal to the cavity axis, respectively.
Without the geometric interpretation as a guide one can consider also
alternative definitions for bright/dark modes -- these do not have
a geometric interpretation but are otherwise equally valid (see Appendix~\ref{subsec:Non-geometric-bright/dark-mode}
for a comparison with the bright/dark mode considered in \cite{shkarin2014optically}).

The couplings in Eq.~\eqref{eq:HamiltonianBD} are given by 
\begin{alignat}{1}
g_{bd}= & \frac{\text{sin}\theta_{\text{tw}}\text{cos}\theta_{\text{tw}}(\omega_{y}^{2}-\omega_{x}^{2})}{2\sqrt{\omega_{b}\omega_{d}}},\label{eq:gbddefm}\\
g_{b}= & g_{x}\sqrt{\frac{\omega_{x}}{\omega_{b}}}\text{sin}\theta_{\text{tw}}+g_{y}\sqrt{\frac{\omega_{y}}{\omega_{b}}}\text{cos}\theta_{\text{tw}},\label{eq:gbdefm}
\end{alignat}
and the frequencies are given by 
\begin{alignat}{1}
\omega_{b}^{2}= & \omega_{x}^{2}\text{sin}^{2}\theta_{\text{tw}}+\omega_{y}^{2}\text{cos}^{2}\theta_{\text{tw}},\label{eq:omegab-1}\\
\omega_{d}^{2}= & \omega_{x}^{2}\text{cos}^{2}\theta_{\text{tw}}+\omega_{y}^{2}\text{sin}^{2}\theta_{\text{tw}}.\label{eq:omegad-1}
\end{alignat}
Importantly, the above derivation of the geometric bright/dark modes
is valid for \emph{any} value of the angle $\theta_{\text{tw}}$ (see
Appendix \ref{subsec:pi/4} for the special case $\theta_{\text{tw}}=\pi/4$
where the expressions simplify further). This is a specific feature
of the coherent scattering setup where one can always decompose the
motions in the transverse tweezer plane into the motion along/perpendicular
to the cavity axis -- the corresponding geometric bright/dark modes
are by construction coupled/decoupled from the cavity mode.

We note that the classical bright/dark \emph{eigenmodes} of the drift
matrix presented in Sec.~\ref{sec:Dark/Bright-modes-for} correspond
closely to $\hat{x}_{b},\hat{x}_{d}$ in the limit $\omega_{x}=\omega_{y}$
where $g_{bd}=0$ and there is no coupling between them. In such a
case $\hat{D}$ can be identified with $\hat{x}_{d}$ while $\hat{B}^{\pm}$
corresponds to the hybridization of $\hat{x}_{b}$ and $\hat{Z}_{L}$.
In general, and in the ``Goldilocks zone'', $g_{bd}\neq0$, and
thus thus the relation between $\hat{B}^{\pm}$, $\hat{D}$ and $\hat{x}_{b}$,
$\hat{x}_{d}$, $\hat{Z}_{L}$ becomes increasingly distorted (as
both $\hat{x}_{b}$, $\hat{x}_{d}$ start to hybridize with $\hat{Z}_{L}$).

In Appendix~\ref{sec:2D-cooling-formulae} we also solve for the
spectra of the hybridised mechanical modes of the system, $\hat{x}_{b}^{\text{2D}}$
and $\hat{x}_{d}^{\text{2D}}$, analogously to Eqs.~\eqref{eq:xF}
and \eqref{eq:yF}, but now given in terms of motions along $\hat{x}_{b}$,
$\hat{x}_{d}$ (and $\hat{Z}_{L}$) rather than motions along the
tweezer axes $\hat{x},\hat{y}$ (and $\hat{Z}_{L}$). The exact closed
form solutions yields completely equivalent heterodyne spectra, but
as we show below decomposing the spectra in terms of $\hat{x}_{b}$,
$\hat{x}_{d}$ mechanical contributions can be less straightforward.

Although the ``Goldilocks zone'' is not in the strongly hybridised
regime, it is still in an intermediate regime where hybridisation
nonetheless plays a critical role. It is instructive to re-examine
the 2D cooling behaviour, but now in terms of the $\hat{x}_{b}^{\text{2D}}$
and $\hat{x}_{d}^{\text{2D}}$ modes. We see from Eq.~\eqref{eq:HamiltonianBD}
that the geometric dark mode, $\hat{x}_{d}$, is decoupled from the
cavity mode -- the optomechanical cooling mechanism must rely on
hybridisation due to the coupling $g_{bd}$ to the bright mode. In
other words, we expect to \emph{sympathetically }cool $\hat{x}_{d}^{\text{2D}}$
only when it is significantly hybridised and contains contributions
from $\hat{x}_{b}$, $\hat{x}_{d}$, and $\hat{Z}_{L}$. We see from
Eq.~(\ref{eq:gbddefm}) that the coupling $g_{bd}$ depends on $\vert\omega_{x}-\omega_{y}\vert$:
only when $\vert\omega_{x}-\omega_{y}\vert$ is large can we can expect
to cool both $\hat{x}_{d}^{\text{2D}}$ and $\hat{x}_{b}^{\text{2D}}$.
In contrast, when $\omega_{x}=\omega_{y}$ the $\hat{x}_{d}^{\text{2D}}$
mode cannot be cooled.

The emergence of the ``Goldilocks zone'' can, in fact also be be
analysed in terms of the 2D optomechanical cooling rates for $\hat{x}_{b}^{\text{2D}}$
and $\hat{x}_{d}^{\text{2D}}$ (see Appendix~\ref{sec:2D-cooling-formulae}
for the derivation): 
\begin{alignat}{1}
\Gamma_{\text{opt,b}} & \equiv\text{Im}\left[2ig_{b}^{2}\eta(\omega_{b})+4g_{bd}^{2}\chi_{d}(\omega_{b})\right],\label{eq:optb}\\
\Gamma_{\text{opt,d}} & \equiv\text{Im}\left[\frac{4g_{bd}^{2}\chi_{b}(\omega_{d})}{1-2ig_{b}^{2}\chi_{b}(\omega_{d})\eta(\omega_{d})}\right],\label{eq:optd}
\end{alignat}
where $\chi_{j}$ is the mechanical susceptibility defined in Eq.~\eqref{eq:mechanical}
with $j=b,d$. In particular, we consider the ideal case $\theta_{\text{tw}}=\pi/4$
where we find simple expressions $\Gamma_{\text{opt,b}}\sim4g_{b}^{2}/\kappa$
and $\Gamma_{\text{opt,d}}\sim g_{bd}^{2}\kappa/g_{b}^{2}$. In order
to cool effectively in 2D both $\Gamma_{\text{opt,b}}$ and $\Gamma_{\text{opt,d}}$
have to be larger than a certain minimum threshold value, $\Gamma_{\text{opt,b,d}}\apprge2\Gamma$
-- these two conditions give rise to the Goldilocks zone. Let us
suppose $g\equiv g_{x}\sim g_{y}$ -- we find $g_{b}\sim\sqrt{2}g$
and $g_{bd}\sim(\omega_{y}-\omega_{x})/2$ -- which further reduces
the 2D optomechanical formulae to $\Gamma_{\text{opt,b}}\sim\frac{8g^{2}}{\kappa}$
and $\Gamma_{\text{opt,d}}\sim\frac{\kappa}{8g^{2}}(\omega_{y}-\omega_{x})^{2}$.
Combining the two constraints we find the condition for the Goldilocks
zone: 
\begin{equation}
\sqrt{\kappa\Gamma/4}\apprle g\apprle\sqrt{\kappa/(16\Gamma)}\vert\omega_{y}-\omega_{x}\vert,\label{eq:boundaries}
\end{equation}
where $\Gamma$ can be loosely identified with the total motional
heating rate. The motional heating rate has a constant contribution
(due to gas collisions) and a power-dependent contribution (from recoil
heating). In first instance we can neglect recoil heating for moderate
powers at the considered pressure of $p\sim10^{-6}\text{mbar}$~\cite{delic2019cavity},
and estimate $\Gamma\sim\gamma n_{B}$, where $\gamma$ is the gas
damping and $n_{B}$ is the mean thermal occupancy. Using the values
in Table \ref{tab:Values-used-in} we estimate for $\Gamma/(2\pi)\sim15\text{kHz}$
and find a qualitative agreement of Eq.~\eqref{eq:boundaries} with
the Goldilocks boundaries shown in Fig.~\ref{fig:phonons_particle}
(the lower and upper dashed green lines).

The phonon occupancy of the modes $\hat{x}_{b}$, $\hat{x}_{d}$ is
however quantitatively different from the one of the modes $\hat{x}$,
$\hat{y}$ -- the two sets of modes have different frequencies ($\omega_{x},\omega_{y}$
versus $\omega_{b},\omega_{d}$) which makes a direct comparison of
the number of phonons difficult. Even disregarding the modest numerical
discrepancies from the frequency differences, estimating the dark
mode phonon occupancy from the heterodyne spectra is a non-trivial
task. To see this we note that the optical field is proportional to
the bright mode (but not to the dark mode): 
\begin{equation}
\hat{a}(\omega)\simeq\eta(\omega)g_{b}\hat{x}_{b}^{\text{2D}}(\omega).\label{eq:bdfield}
\end{equation}
One can then directly extract the bright mode displacement PSD from
the normalised heterodyne PSD:
\begin{equation}
\frac{S_{\text{het}}(\omega)}{g_{b}^{2}|\eta(\omega)|^{2}}\simeq S_{x_{b}x_{b}}(\omega),\label{eq:afield2}
\end{equation}
where $S_{x_{b}x_{b}}(\omega)=\langle|\hat{x}_{b}^{\text{2D}}(\omega)|\rangle$.
We can thus obtain an occupancy $n_{b}$ by integrating the area under
the heterodyne PSD rescaled by the factor $g_{b}^{2}|\eta(\omega)|^{2}$
(see Eq.~\eqref{Phon} with $j=x_{b},x_{d}$ and define $n_{j}\equiv n_{x_{j}}$).
On the other hand, from the measured heterodyne spectra one is not
able to measure $S_{x_{d}x_{d}}(\omega)=\langle|\hat{x}_{d}^{\text{2D}}(\omega)|\rangle$
and thus one cannot directly obtain an estimate for the corresponding
phonon occupancy $n_{d}$. However, by comparing Eqs.~\eqref{eq:afield}
and \eqref{eq:afield2} (and using $g_{b}\sim\sqrt{2}g$ at $\theta_{\text{tw}}=\pi/4$)
we find $S_{x_{b}x_{b}}(\omega)\sim(S_{xx}(\omega)+S_{yy}(\omega))/2$
and thus $n_{b}\sim(n_{x}+n_{y})/2$. When both the bright/dark mode
are cooled with the same optomechanical rate (i.e., $\Gamma_{\text{opt,b}}\sim\Gamma_{\text{opt,d}}$)
one has $n_{b}\sim n_{d}$, and one can indirectly infer that the
area under $S_{x_{d}x_{d}}(\omega)=\langle|\hat{x}_{d}^{\text{2D}}(\omega)|\rangle$
will give $n_{d}\sim(n_{x}+n_{y})/2$. We thus find $n_{x}+n_{y}\sim n_{b}+n_{d}$.
This explains why both the tweezer-based and the cavity-based modes
approximately agree about the total phonon occupancy and lead to roughly
the same Goldilocks zone.

\section{Discussion\label{sec:Discussion}}

A previous theoretical study that investigated hybridisation due to
optomechanical interaction via coherent scattering by levitated nanoparticles,
left an important gap in understanding: that study \citep{torovs2020quantum}
investigated the quasi-1D dynamics arising for trapping at a node,
relevant to recent quantum ground-state cooling experiments. However,
the quasi-1D behavior $g_{x}\gg g_{y}$ of one strong-coupled mode,
hybridising with one weak coupled mode, obtained for $\theta_{tw}\to\pi/2$,
differs fundamentally from the 2D scenario of two strong-coupled modes
that arises as $\theta_{tw}\to\pi/4$, where $g_{x}\simeq g_{y}$.

Here we have shown that simultaneous quantum ground-state cooling
of both strong coupled modes is no longer achievable simply by the
usual 1D strategy of increasing the coupling strengths/cooperativities
of the individual modes: one must also factor in the essentially 2D
phenomenon of the formation of dark modes that decouple from the optical
field, as well as other requirements.

We have investigated the levitated nanoparticle motion in the tweezer
transverse ($x$-$y$) plane with an optical cavity. We showed that
efficient cooling of the $\hat{x}$ and $\hat{y}$ motion to their
quantum ground state must obey certain constraints relating the difference
of mechanical frequencies $|\omega_{x}-\omega_{y}|$ to the coupling
rates $g_{x},g_{y}$ as well as the cavity decay rate $\kappa$. We
found that cooling and standard thermometry are efficient for a sufficiently
elliptical optical trap, while for a more spherical trap the cooling
will be hindered by strong three-way mode hybridization with the cavity
mode. We found also the optimal particle size that satisfies the conditions,
thus allowing for 2D ground state cooling in the current experimental
setup.

The analysis of the 2D levitated nanoparticle problem also found that
the dark/bright modes have a geometric interpretation in terms of
rotations in the $x$-$y$ plane. Importantly, the transformation
is not a trivial rotation because of the non-equivalence of the $\hat{x},\hat{y}$
phononic modes (we assume trap ellipticity) so the modification for
$\omega_{x}\neq\omega_{y}$ is discussed. In addition, we also considered
in the calculations the effects of scattering force so as to be able
to reliably simulate larger particles.

Free-fall experiments that propose recycling of particles -- where
particles would be trapped again after a sufficiently long free-fall
time -- require the nanoparticle energy to be low in all three translational
motions. The motion along the optical tweezer axis can be cooled to
its ground state via feedback cooling~\citep{tebbenjohanns2020motional, magrini2020optimal},
thus extending our 2D scheme to fully prepare nanoparticles for free-fall
experiments. In addition, the uncoupled three degrees of freedom can
be used as a (quantum) sensor of forces acting along a specific direction,
such as terrestrial gravity fluctuations~\citep{harms2019terrestrial}.

\section*{Acknowledgments}

This work was presented at the ICTP workshop Frontiers of Nanomechanics
(September 2020) and at the conference Quantum Nanophotonics (March 2021). We are extremely grateful to Vladan Vuleti\'{c}
for insightful discussions about bright/dark modes. 
This work was supported by the Engineering and Physical Sciences Research Council [EP/N031105/1  and  EP/L015242/1]. MT acknowledges funding by the Leverhulme Trust (RPG-2020-197). UD acknowledges support from the European Research Council (ERC CoG QLev4G) and the research platform TURIS at the University of Vienna.

{\em Following submission of the present study, we became aware
of a new interesting experimental study by Ranfagni et al arXiv:2012.15265
which, for $\theta_{tw}\simeq0.4\pi$ exhibits a scenario intermediate
between the quasi 1D limit and the 2D regimes we investigate here.}

\appendix

\section{Notes on scattering force\label{sec:Notes-on-scattering}}

In this appendix we look at the modification of the optomechanical
interaction due to the shifted equilibrium position of the nanoparticle
along the z-axis (with respect to the tweezer trap center). In particular,
such a shifted equilibrium arises from the scattering radiation pressure
force as we increase the particle size. We first find the new equilibrium
position (Sec.~\ref{subsec:z-axis-equilibrium-position}) and then
calculate the new optomechanical couplings (Sec.~\ref{subsec:Interaction-potential}).
We finally show that by using appropriately rotated optical quadratures
the optomechanical formulae derived for the case of negligible scattering
force -- appropriate for small nanoparticles -- extend also to the
case with with a shifted equilibrium position (Sec.~\ref{subsec:Rotated-optical-quadratures}).

\subsection{z--axis equilibrium position\label{subsec:z-axis-equilibrium-position}}

The competition between the gradient force, $F_{\text{grad}}$, and
the scattering force, $F_{\text{scatt}}$, modifies the nanoparticle's
equilibrium position, $z_{\ensuremath{0}}$, along the z-axis (the
tweezer symmetry axis). In particular, the gradient and scattering
force are given by~\citep{harada1996radiation}:

\begin{alignat}{1}
F_{\text{grad}} & =-\frac{2\pi R^{3}}{c}\frac{\epsilon_{R}-1}{\epsilon_{R}+2}\partial_{z}I(z),\\
F_{\text{scatt}} & =\frac{8\pi k^{4}R^{6}}{3c}\left(\frac{\epsilon_{R}-1}{\epsilon_{R}+2}\right)^{2}I(z),
\end{alignat}
respectively, where $\epsilon_{R}$ is the relative dielectic permittivity,
$c$ is the speed of light, $k=\frac{2\pi}{\lambda}$, $\lambda$
is the wavelength, and $R$ is the particle radius. The laser intensity
along the tweezer axis is given by 
\begin{equation}
I(z)\equiv\frac{2P_{\text{tw}}}{\pi w_{x}w_{y}\left[1+(\frac{z}{z_{R}})^{2}\right]},
\end{equation}
where $P_{\text{tw}}$ is the laser power at the center of the trap,
$w_{x}$, $w_{y}$ are the beam waists, and $z_{R}=\pi w_{x}w_{y}/\lambda$
is the Rayleigh range. We readily find the equilibrium position, $z_{0}$,
from the condition $F_{\text{grad}}+F_{\text{scatt}}=0$. Assuming
$z_{0}/z_{R}\ll1$ we find a simple result:

\begin{equation}
z_{0}=\frac{\epsilon_{R}-1}{\epsilon_{R}+2}\frac{2k^{4}z_{R}^{2}}{3}R^{3},\label{eq:z0}
\end{equation}
but can otherwise numerically solve for the equilibrium position.
Importantly, the larger the particle radius, $R$, the more the equilibrium
position, $z_{0}$, is displaced from the Gaussian beam focus~\citep{timberlake2019static}.

\subsection{Optomechanical couplings\label{subsec:Interaction-potential}}

We start from the coherent scattering interaction potential~\citep{delic2019cavity}:

\begin{alignat}{1}
\frac{\hat{V}_{\text{int}}}{\hbar}= & -E_{d}\text{cos}\left[\phi+k(\hat{x}\text{sin}(\theta_{\text{tw}})+\hat{y}\text{cos}(\theta_{\text{tw}}))\right]\nonumber \\
 & \qquad\times\left(\hat{a}e^{-i\xi}+\hat{a}^{\dagger}e^{i\xi}\right),\label{eq:potential}
\end{alignat}
where $\xi=kz+\Phi(z)$, and $\Phi(z)=-\text{arctan}(z/z_{R})$ is
the Gouy phase. The equilibrium position of the nanoparticle with
respect to the tweezer trap center will be denoted by $(x_{0},y_{0},z_{0})$,
where we assume $x_{0}=y_{0}=0$, while $z_{0}$ is given by Eq.~\ref{eq:z0}
(for $z_{0}/z_{R}\ll1$) or obtained by solving numerically for the
equilibrium position.

Expanding the interaction potential in Eq.~\eqref{eq:potential}
to quadratic order we find the following couplings:

\begin{alignat}{1}
\frac{\hat{V}_{\text{int}}}{\hbar}= & g_{xZ}^{\xi}\hat{x}\hat{Z}+g_{xP}^{\xi}\hat{x}\hat{P}+g_{yZ}^{\xi}\hat{y}\hat{Z}+g_{yP}^{\xi}\hat{y}\hat{P}+g_{zZ}^{\xi}\hat{z}\hat{Z}\nonumber \\
 & +g_{zP}^{\xi}\hat{z}\hat{P}+g_{xy}^{\xi}\hat{x}\hat{y}+g_{xz}^{\xi}\hat{x}\hat{z}+g_{yz}^{\xi}\hat{y}\hat{z}.\label{eq:potential2}
\end{alignat}
Assuming $kz_{R}\gg1$, neglecting terms of order $\mathcal{O}(z_{0}/z_{R})$,
while still retaining the phase $\xi\sim kz_{0}$, we find simple
couplings

\begin{alignat}{2}
g_{xZ}^{\xi}= & g_{xZ}\text{cos}(\xi), & g_{xP}^{\xi}= & g_{xZ}\text{sin}(\xi),\\
g_{yZ}^{\xi}= & g_{yZ}\text{cos}(\xi), & g_{yZ}^{\xi}= & g_{yZ}\text{sin}(\xi),\\
g_{xy}^{\xi}= & \frac{g_{xy}}{Y_{0}}(Z_{0}\text{cos}(\xi)+P_{0}\text{sin}(\xi)).
\end{alignat}
where

\begin{alignat}{1}
g_{xZ}= & E_{d}k\text{sin}(\theta_{\text{tw}})\text{sin}(\phi)x_{\text{zpf}},\\
g_{yZ}= & E_{d}k\text{cos}(\theta_{\text{tw}})\text{sin}(\phi)y_{\text{zpf}},\\
g_{xy}= & E_{d}k^{2}Z_{0}\text{sin}(\theta_{\text{tw}})\text{cos}(\theta_{\text{tw}})\text{cos}(\phi)x_{\text{zpf}}y_{\text{zpf}},
\end{alignat}
and $Z_{0}$,$P_{0}$ denote the mean values of the optical quadratures.

For completeness we list also the z-couplings:

\begin{alignat}{1}
g_{zZ}^{\xi} & =-g_{zP}\text{sin}(\xi),\qquad g_{zP}^{\xi}=g_{zP}\text{cos}(\xi),\\
g_{xz}^{\xi} & =\frac{g_{xz}}{P_{0}}\left[P_{0}\text{cos}(\xi)-Z_{0}\text{sin}(\xi)\right],\\
g_{yz}^{\xi} & =\frac{g_{yz}}{P_{0}}\left[P_{0}\text{cos}(\xi)-Z_{0}\text{sin}(\xi)\right],
\end{alignat}
where

\begin{alignat}{1}
g_{zP} & =-E_{d}k\text{cos}(\phi)z_{\text{zpf}},\\
g_{xz} & =E_{d}k^{2}P_{0}\text{sin}(\theta_{\text{tw}})\text{sin}(\phi)x_{\text{zpf}}z_{\text{zpf}},\\
g_{yz} & =E_{d}k^{2}P_{0}\text{cos}(\theta_{\text{tw}})\text{sin}(\phi)y_{\text{zpf}}z_{\text{zpf}}.
\end{alignat}
The couplings $g_{xZ}$,$g_{yZ}$,$g_{xy}$,$g_{zP}$,$g_{xz}$,$g_{yz}$
have been previously obtained by neglecting the scattering force and
are valid for small nanoparticles~\citep{torovs2020quantum}.

\subsection{Rotated optical quadratures\label{subsec:Rotated-optical-quadratures}}

It is instructive to compare the case with negligible scattering force
(i.e., $z_{0}=0$ and $\xi=0$) with the case of an arbitrary z-axis
displacement from the tweezer trap center (i.e., $z_{0}>0$ and thus
$\xi>0$). In particular, we introduce the rotated optical quadratures:

\begin{alignat}{1}
\left[\begin{array}{c}
\hat{Z}^{\xi}\\
\hat{P}^{\xi}
\end{array}\right] & =\left[\begin{array}{cc}
\text{cos}(\xi) & -\text{sin}(\xi)\\
\text{sin}(\xi) & \text{cos}(\xi)
\end{array}\right]\left[\begin{array}{c}
\hat{Z}\\
\hat{P}
\end{array}\right],\label{eq:optrotation}
\end{alignat}
where the angle of rotation is $\xi\sim kz_{0}$, and $\hat{Z}^{\xi}$,$\hat{P}^{\xi}$
($\hat{Z}$, $\hat{P}$) denote the optical quadratures in the case
with (without) the $z_{0}$ displacement.

Let us first consider the mean values. From Eq.~\eqref{eq:potential},
writing the corresponding classical equations of motion, we find that
the mean-value of the optical quadratures are given by

\begin{alignat}{1}
Z_{0}^{\xi} & =-\frac{E_{d}\text{cos}(\phi)}{\Delta^{2}+(\frac{\kappa}{2})^{2}}\left[2\Delta\text{cos}(\xi)-\kappa\text{sin}(\xi)\right],\label{eq:Y0xi}\\
P_{0}^{\xi} & =-\frac{E_{d}\text{cos}(\phi)}{\Delta^{2}+(\frac{\kappa}{2})^{2}}\left[2\Delta\text{sin}(\xi)+\kappa\text{cos}(\xi)\right].\label{eq:P0xi}
\end{alignat}
where $\Delta$ is the detuning, and $\kappa$ the cavity decay rate.
If we set $z_{0}=0$ (and hence $\xi=0$) we find the simplified expression
for the amplitude and phase quadratures, which we denote by $Z_{0}$
and $P_{0}$, respectively. From Eqs.~\eqref{eq:Y0xi} and \eqref{eq:P0xi}
we readily see that $Z_{0}^{\xi}$, $P_{0}^{\xi}$ and $Z_{0}$, $P_{0}$
are related by the rotation introduced in Eq.~\eqref{eq:optrotation}.

Using now the rotated quadratures,$\hat{Z}^{\xi}$,$\hat{P}^{\xi}$,
and the rotated mean values, $Z_{0}^{\xi}$, $P_{0}^{\xi}$, the interaction
potential in Eq.~\eqref{eq:potential2} reduces to the expression:

\begin{alignat}{1}
\frac{\hat{V}_{\text{int}}}{\hbar}= & +g_{xZ}\hat{x}\hat{Z}^{\xi}+g_{yZ}\hat{y}\hat{Z}^{\xi}+g_{zP}\hat{z}\hat{P}^{\xi}\nonumber \\
 & +g_{xy}^{\xi}\hat{x}\hat{y}+g_{xz}^{\xi}\hat{x}\hat{z}+g_{yz}^{\xi}\hat{y}\hat{z},\label{eq:potential3}
\end{alignat}
where

\begin{alignat}{1}
g_{xy}^{\xi} & =E_{d}k^{2}Z_{0}^{\xi}\text{sin}(\theta_{\text{tw}})\text{cos}(\theta_{\text{tw}})\text{cos}(\phi)x_{\text{zpf}}y_{\text{zpf}},\label{eq:gxyxi}\\
g_{xz}^{\xi} & =E_{d}k^{2}P_{0}^{\xi}\text{sin}(\theta_{\text{tw}})\text{sin}(\phi)x_{\text{zpf}}z_{\text{zpf}},\\
g_{yz}^{\xi} & =E_{d}k^{2}P_{0}^{\xi}\text{cos}(\theta_{\text{tw}})\text{sin}(\phi)y_{\text{zpf}}z_{\text{zpf}}.\label{eq:gyzxi}
\end{alignat}
We note that the potential in Eq.~\eqref{eq:potential3} has the
same form of the potential previously obtained for the case of small
nanoparticles~\citep{torovs2020quantum} (i.e. where one can neglect
the displacement, $z_{0}$, due to the scattering force): one formally
replaces $\hat{Z}\rightarrow\hat{Z}^{\xi}$ and $\hat{P}\rightarrow\hat{P}^{\xi}$.
Thus all the formulae obtained in \citep{torovs2020quantum} remain
valid also when we consider a significant non-zero displacement along
the z-axis (displacement from the tweezer trap center), provided we
use the rotated mean values, $Z_{0}^{\xi}$ and $P_{0}^{\xi}$, given
in Eqs.~\eqref{eq:Y0xi} and \eqref{eq:P0xi}, respectively.

For the special case considered in the main text the interaction potential
remains of the same form as in the case without any z-axis displacement:

\begin{alignat}{1}
\frac{\hat{V}_{\text{int}}}{\hbar} & =g_{x}\hat{x}\hat{Z}_{L}+g_{y}\hat{y}\hat{Z}_{L},\label{eq:potential4}
\end{alignat}
where we have defined $\hat{Z}_{L}\equiv\hat{Z}^{\xi}$, $g_{x}\equiv g_{xZ}$,
and $g_{y}\equiv g_{yZ}$.

\section{Bright/dark modes and geometric rotations}

In this section we introduce bright/dark modes and show that they have a simple geometric interpretation as the motion along/perpendicular to the cavity axis (in the transverse tweezer plane). We first recall the usual 2D matrix that rotates the reference frames -- we show how it induces a rotation of the physical positions and momenta while it leads to a distorted transformation of the associated canonical position and momenta (Sec.~\ref{sec:2dRotations}). We then proceed to define the geometric bright/dark modes -- we obtain the coupling of the bright mode to the optical field as well as the induced coupling to the dark mode (Sec.~\ref{sec:brightdark}) and provide the simplified geometric bright/dark mode couplings at $\theta_{\text{tw}}=\pi/4$ (Sec.~\ref{subsec:pi/4}). For completeness we also compare the geometric bright/dark modes with the alternative bright/dark modes used in Ref.~\cite{shkarin2014optically}  (Sec.~\ref{subsec:Non-geometric-bright/dark-mode}). Finally, we derive the 2D optomechanical cooling formulae for the bright/dark modes (Sec.~\ref{sec:2D-cooling-formulae}).

\subsection{Geometric and canonical rotations\label{sec:2dRotations}}

The geometric rotation introduced in Eq.~\eqref{eq:rotation} relates
the coordinates of the ``tweezer reference frame'' with the coordinates
of the ``cavity reference frame'' (see Fig.~\ref{Fig1}). Such
a rotation induces a transformation of the position and momenta --
we will label physical positions (momenta) with capital letters( $\hat{X},\hat{Y}$($\hat{P}_{X},\hat{P}_{Y}$)
for the tweezer frame and $\hat{X}^{\text{(c)}},\hat{Y}^{\text{(c)}}$($\hat{P}_{X}^{\text{(c)}},\hat{P}_{Y}^{\text{(c)}}$)
for the cavity frame). Specifically, the physical positions transform
as

\begin{equation}
\left[\begin{array}{c}
\hat{X}^{\text{(c)}}\\
\hat{Y}^{\text{(c)}}
\end{array}\right]=\left[\begin{array}{cc}
\text{sin}\theta_{\text{tw}} & \text{cos}\theta_{\text{tw}}\\
-\text{cos}\theta_{\text{tw}} & \text{sin}\theta_{\text{tw}}
\end{array}\right]\left[\begin{array}{c}
\hat{X}\\
\hat{Y}
\end{array}\right],\label{eq:rotationXX}
\end{equation}
and the physical momenta transform as

\begin{equation}
\left[\begin{array}{c}
\hat{P}_{X}^{\text{(c)}}\\
\hat{P}_{Y}^{\text{(c)}}
\end{array}\right]=\left[\begin{array}{cc}
\text{sin}\theta_{\text{tw}} & \text{cos}\theta_{\text{tw}}\\
-\text{cos}\theta_{\text{tw}} & \text{sin}\theta_{\text{tw}}
\end{array}\right]\left[\begin{array}{c}
\hat{P}_{X}\\
\hat{P}_{Y}
\end{array}\right].\label{eq:rotationPP}
\end{equation}
We recall that the physical positions/momenta are related to the canonical
positions/momenta (denoted by lowercase letters) through a simple
multiplicative rescaling:

\begin{alignat}{2}
\hat{X} & =x_{\text{zpf}}\hat{x},\qquad & \hat{Y} & =y_{\text{zpf}}\hat{y},\\
\hat{P}_{X} & =p_{x,\text{zpf}}\,\hat{p}_{x}, & \hat{P}_{Y} & =p_{y,\text{zpf}}\,\hat{p}_{y},\\
\hat{X}^{\text{(c)}} & =x_{\text{zpf}}^{\text{(c)}}\hat{x}^{\text{(c)}},\qquad & \hat{Y} & =y_{\text{zpf}}^{\text{(c)}}\hat{y}^{\text{(c)}},\label{eq:XC}\\
\hat{P}_{X}^{\text{(c)}} & =p_{x,\text{zpf}}^{\text{(c)}}\,\hat{p}_{x}^{\text{(c)}}, & \hat{P}_{Y} & =p_{y,\text{zpf}}^{\text{(c)}}\,\hat{p}_{y}^{\text{(c)}},\label{PC}
\end{alignat}
where the zero-point motions are given by 
\begin{alignat}{2}
x_{\text{zpf}} & =\sqrt{\frac{\hbar}{2m\omega_{x}}},\qquad & y_{\text{zpf}} & =\sqrt{\frac{\hbar}{2m\omega_{y}}},\\
p_{x,\text{zpf}} & =\sqrt{\frac{\hbar m\omega_{x}}{2}}, & p_{y,\text{zpf}} & =\sqrt{\frac{\hbar m\omega_{y}}{2}},\\
x_{\text{zpf}}^{\text{(c)}} & =\sqrt{\frac{\hbar}{2m\omega_{x}^{\text{(c)}}}},\qquad & y_{\text{zpf}}^{\text{(c)}} & =\sqrt{\frac{\hbar}{2m\omega_{y}^{\text{(c)}}}},\\
p_{x,\text{zpf}}^{\text{(c)}} & =\sqrt{\frac{\hbar m\omega_{x}^{\text{(c)}}}{2}}, & p_{y,\text{zpf}}^{\text{(c)}} & =\sqrt{\frac{\hbar m\omega_{y}^{\text{(c)}}}{2}},
\end{alignat}
where $\omega_{x}$, $\omega_{y}$ ( $\omega_{x}^{\text{(c)}}$,$\omega_{y}^{\text{(c)}}$)
are the oscillation frequencies along the axis of the tweezer frame
(cavity frame). Hence the rotations in Eqs.~\eqref{eq:rotationXX}
and \eqref{eq:rotationPP} can be recast in terms of the canonical
positions and momenta:

\begin{alignat}{1}
\left[\begin{array}{c}
\hat{x}^{\text{(c)}}\\
\hat{y}^{\text{(c)}}
\end{array}\right] & =\left[\begin{array}{cc}
\sqrt{\frac{\omega_{x}^{\text{(c)}}}{\omega_{x}}}\text{sin}\theta_{\text{tw}} & \sqrt{\frac{\omega_{x}^{\text{(c)}}}{\omega_{y}}}\text{cos}\theta_{\text{tw}}\\
-\sqrt{\frac{\omega_{y}^{\text{(c)}}}{\omega_{x}}}\text{cos}\theta_{\text{tw}} & \sqrt{\frac{\omega_{y}^{\text{(c)}}}{\omega_{y}}}\text{sin}\theta_{\text{tw}}
\end{array}\right]\left[\begin{array}{c}
\hat{x}\\
\hat{y}
\end{array}\right],\label{eq:rotationXXC}
\end{alignat}
and

\begin{equation}
\left[\begin{array}{c}
\hat{p}_{x}^{\text{(c)}}\\
\hat{p}_{y}^{\text{(c)}}
\end{array}\right]=\left[\begin{array}{cc}
\sqrt{\frac{\omega_{x}}{\omega_{x}^{\text{(c)}}}}\text{sin}\theta_{\text{tw}} & \sqrt{\frac{\omega_{y}}{\omega_{x}^{\text{(c)}}}}\text{cos}\theta_{\text{tw}}\\
-\sqrt{\frac{\omega_{x}}{\omega_{y}^{\text{(c)}}}}\text{cos}\theta_{\text{tw}} & \sqrt{\frac{\omega_{y}}{\omega_{y}^{\text{(c)}}}}\text{sin}\theta_{\text{tw}}
\end{array}\right]\left[\begin{array}{c}
\hat{p}_{x}\\
\hat{p}_{y}
\end{array}\right],\label{eq:rotationPPC}
\end{equation}
respectively. We thus see that a geometric rotation of the coordinates
introduces a distorted transformation for the canonical variables.
Only when the optical trap is perfectly degenerate (i.e, $\omega_{x}=\omega_{y}$
which implies $\omega_{x}=\omega_{y}=\omega_{x}^{\text{(c)}}=\omega_{y}^{\text{(c)}}$)
we find that the transformations of the canonical variables in Eqs.~\eqref{eq:rotationXXC}
and \eqref{eq:rotationPPC} reduce to geometric rotations.

\subsection{Definition of bright/dark modes\label{sec:brightdark}}

We start from the Hamiltonian in Eq.~\eqref{eq:2DHamiltonian} which
we recast in terms of the physical positions/momenta (see previous
Sec.~\ref{sec:2dRotations}):

\begin{alignat}{1}
\hat{H} & =\frac{1}{2m}\left[\hat{P}_{X}^{2}+\hat{P}_{Y}^{2}\right]-\frac{\Delta}{4}(\hat{Z}_{L}^{2}+\hat{P}_{L}^{2})\nonumber \\
 & +\frac{m\omega_{x}^{2}}{2}\hat{X}^{2}+\frac{m\omega_{y}^{2}}{2}\hat{Y}^{2}+\left(\frac{\hbar g_{x}}{x_{\text{zpf}}}\hat{X}+\frac{\hbar g_{y}}{y_{\text{zpf}}}\hat{Y}\right)\hat{Z}_{L}.\label{eq:Hp}
\end{alignat}
We first note that the terms on the first line of Eq.~\eqref{eq:Hp}
remain of the same form under the action of a geometric rotation and
will thus be omitted in the following analysis (using Eq.~\eqref{eq:rotationPP}
one can readily show that $\hat{P}_{X}^{2}+\hat{P}_{Y}^{2}$ changes
to $\hat{P}_{X}^{\text{(c)}2}+\hat{P}_{Y}^{\text{(c)}2}$, while the
term $\hat{Z}_{L}^{2}+\hat{P}_{L}^{2}$ is not affected).

We now rewrite Eqs.~\eqref{eq:Hp} in terms of the transformed positions/momenta
(using the inverse transformation of Eq.~\eqref{eq:rotationXX}):

\begin{alignat}{1}
\hat{H} & =\frac{m}{2}\left(\omega_{x}^{2}\text{sin}^{2}\theta_{\text{tw}}+\omega_{y}^{2}\text{cos}^{2}\theta_{\text{tw}}\right)\hat{X}^{\text{(c)}2}\nonumber \\
 & +\frac{m}{2}\left(\omega_{x}^{2}\text{cos}^{2}\theta_{\text{tw}}+\omega_{y}^{2}\text{sin}^{2}\theta_{\text{tw}}\right)\hat{Y}^{\text{(c)}2}\nonumber \\
 & +m\left(\text{sin}\theta_{\text{tw}}\text{cos}\theta_{\text{tw}}(\omega_{y}^{2}-\omega_{x}^{2})\right)\hat{X}^{\text{(c)}}\hat{Y}^{\text{(c)}}\nonumber \\
 & +\left[\frac{\hbar g_{x}}{x_{\text{zpf}}}\text{sin}\theta_{\text{tw}}+\frac{\hbar g_{y}}{y_{\text{zpf}}}\text{cos}\theta_{\text{tw}}\right]\hat{X}^{\text{(c)}}\hat{Z}_{L}\nonumber \\
 & +\left[-\frac{\hbar g_{x}}{x_{\text{zpf}}}\text{cos}\theta_{\text{tw}}+\frac{\hbar g_{y}}{y_{\text{zpf}}}\text{sin}\theta_{\text{tw}}\right]\hat{Y}^{\text{(c)}}\hat{Z}_{L}.\label{eq:long}
\end{alignat}
We now make the key observation -- the coupling between $\hat{Y}^{\text{(c)}}$
and $\hat{Z}_{L}$ in the last line of Eq.~\eqref{eq:long} vanishes
(we recall from Eq.~\eqref{eq:couplings} that $g_{x}\sim\text{sin}\theta_{\text{tw}}$$/x_{\text{zpf}}$
and $g_{y}\sim\text{cos}\theta_{\text{tw}}$$/y_{\text{zpf}}$) whilst
the coupling between $\hat{Y}^{\text{(c)}}$ and $\hat{Z}_{L}$ reduces
to $-\hbar E_{d}k$ (the driving amplitude $E_{d}$ and the wave-vector
$k$ are defined below Eq.~\eqref{eq:couplings}). It is thus appropriate
to identify $\hat{Y}^{\text{(c)}}$as the dark mode (motion orthogonal
to the cavity axis), and $\hat{X}^{\text{(c)}}$as the bright mode
(motion parallel to the cavity axis)-- namely, the \emph{geometric}
bright/dark modes.

In summary, we have found that the geometric rotation of the reference
frame (with the angle of rotation matching the angle between the tweezer
polarization and cavity symmetry axis) leads to the bright/dark mode.
We thus re-label the mechanical modes in the cavity reference frame
as:

\begin{alignat}{1}
\hat{X}^{\text{(c)}},\hat{Y}^{\text{(c)}},\hat{x}^{\text{(c)}},\hat{y}^{\text{(c)}} & \rightarrow\hat{X}_{b},\hat{X}_{d},\hat{x}_{b},\hat{x}_{d},\label{eq:XC-1}\\
\hat{P}_{X}^{\text{(c)}},\hat{P}_{Y}^{\text{(c)}},\hat{p}_{x}^{\text{(c)}},\hat{p}_{y}^{\text{(c)}} & \rightarrow\hat{P}_{b},\hat{P}_{d},\hat{p}_{b},\hat{p}_{d}.\label{PC-1}
\end{alignat}
From the first two lines of Eq.~\eqref{eq:long} we can read the
transformed mechanical frequencies:

\begin{alignat}{1}
\omega_{b}^{2}\equiv(\omega_{x}^{\text{(c)}})^{2}= & \omega_{x}^{2}\text{sin}^{2}\theta_{\text{tw}}+\omega_{y}^{2}\text{cos}^{2}\theta_{\text{tw}},\label{eq:omegab}\\
\omega_{d}^{2}\equiv(\omega_{y}^{\text{(c)}})^{2}= & \omega_{x}^{2}\text{cos}^{2}\theta_{\text{tw}}+\omega_{y}^{2}\text{sin}^{2}\theta_{\text{tw}}.\label{eq:omegad}
\end{alignat}
Furthermore, we can then rewrite the interaction terms of Eq.~\eqref{eq:long}
(third and fourth lines) as:

\begin{alignat}{1}
\frac{\hat{H}}{\hbar} & =g_{bd}\hat{x}_{b}\hat{x}_{d}+g_{b}\hat{x}_{b}\hat{Z}_{L},\label{eq:bdh}
\end{alignat}
where 
\begin{alignat}{1}
g_{bd}= & \frac{\text{sin}\theta_{\text{tw}}\text{cos}\theta_{\text{tw}}(\omega_{y}^{2}-\omega_{x}^{2})}{2\sqrt{\omega_{b}\omega_{d}}},\label{eq:gbddef}\\
g_{b}= & g_{x}\sqrt{\frac{\omega_{x}}{\omega_{b}}}\text{sin}\theta_{\text{tw}}+g_{y}\sqrt{\frac{\omega_{y}}{\omega_{b}}}\text{cos}\theta_{\text{tw}}.\label{eq:gbdef}
\end{alignat}
Using Eq.~\eqref{eq:couplings} we can rewrite the coupling in Eq.~\eqref{eq:gbdef}
as $g_{b}=-E_{d}k\sqrt{\frac{\hbar}{2m\omega_{b}}}$ (the driving
amplitude $E_{d}$ and the wave-vector $k$ are defined below Eq.~\eqref{eq:couplings}).

\subsection{Dark/bright mode couplings at \textmd{\normalsize{}{}$\theta_{\text{tw}}=\pi/4$
\label{subsec:pi/4}}}

The analysis in Appendix~\ref{sec:2dRotations} and \ref{sec:brightdark}
is valid for any angle $\theta_{\text{tw}}$. We now write the couplings
for the special case $\theta_{\text{tw}}=\pi/4$ considered in the
main text where one has $\text{sin}\theta_{\text{tw}}=\text{cos}\theta_{\text{tw}}=1/\sqrt{2}$.
From Eqs.~\eqref{eq:omegab} and \eqref{eq:omegad} we first note
that $\omega_{b}=\omega_{d}$ and define 
\begin{equation}
\omega_{bd}^{2}\equiv\omega_{b}^{2}=\omega_{d}^{2}=\frac{\omega_{x}^{2}+\omega_{y}^{2}}{2}.\label{eq:omegabd}
\end{equation}
The couplings in Eqs.~\eqref{eq:gbddef} and \eqref{eq:gbdef} simplify
to

\begin{alignat}{1}
g_{bd}= & \frac{\omega_{y}^{2}-\omega_{x}^{2}}{4\omega_{bd}},\qquad g_{b}=g_{x}\sqrt{\frac{\omega_{x}}{2\omega_{bd}}}+g_{y}\sqrt{\frac{\omega_{y}}{2\omega_{bd}}},
\end{alignat}
respectively.

\subsection{Non-geometric bright/dark mode\label{subsec:Non-geometric-bright/dark-mode}}

It is instructive to compare the geometric bright/dark modes introduced
above with the alternative bright/dark modes introduced in \cite{shkarin2014optically}
which are obtained by rotations of the abstract space spanned by the
canonical positions and momenta -- we will refer to the latter as
the non-geometric bright/dark modes. Specifically, in place of Eqs.~\eqref{eq:rotationXXC}
and ~\eqref{eq:rotationPPC} we now consider a simple rotation of
the canonical positions,

\begin{alignat}{1}
\left[\begin{array}{c}
\hat{x}^{\text{(c)}}\\
\hat{y}^{\text{(c)}}
\end{array}\right] & =\left[\begin{array}{cc}
\text{sin}\theta_{\text{ng}} & \text{cos}\theta_{\text{ng}}\\
-\text{cos}\theta_{\text{ng}} & \text{sin}\theta_{\text{ng}}
\end{array}\right]\left[\begin{array}{c}
\hat{x}\\
\hat{y}
\end{array}\right],\label{eq:rotationC1}
\end{alignat}
and of the canonical momenta,

\begin{alignat}{1}
\left[\begin{array}{c}
\hat{p}_{x}^{\text{(c)}}\\
\hat{p}_{y}^{\text{(c)}}
\end{array}\right] & =\left[\begin{array}{cc}
\text{sin}\theta_{\text{ng}} & \text{cos}\theta_{\text{ng}}\\
-\text{cos}\theta_{\text{ng}} & \text{sin}\theta_{\text{ng}}
\end{array}\right]\left[\begin{array}{c}
\hat{p}_{x}\\
\hat{p}_{y}
\end{array}\right],\label{eq:rotationC2}
\end{alignat}
where $\theta_{\text{ng}}$ is the angle of the abstract rotation.

We now express Eq.~\eqref{eq:2DHamiltonian} in the transformed coordinates
to find:

\begin{alignat}{1}
\frac{\hat{H}}{\hbar} & =\frac{1}{4}(\omega_{x}\text{sin}^{2}\theta_{\text{ng}}+\omega_{y}\text{cos}^{2}\theta_{\text{ng}})\left[\hat{x}^{\text{(c)}2}+\hat{p}_{x}^{\text{(c)}2}\right]\nonumber \\
 & +\frac{1}{4}(\omega_{x}\text{sin}^{2}\theta_{\text{ng}}+\omega_{y}\text{cos}^{2}\theta_{\text{ng}})\left[\hat{y}^{\text{(c)}2}+\hat{p}_{y}^{\text{(c)}2}\right]\nonumber \\
 & +\frac{1}{2}(\omega_{y}-\omega_{x})\text{sin}\theta_{\text{ng}}\text{cos}\theta_{\text{ng}}\left[\hat{x}^{\text{(c)}}\hat{y}^{\text{(c)}}+\hat{p}_{x}^{\text{(c)}}\hat{p}_{y}^{\text{(c)}}\right]\nonumber \\
 & +\left[g_{x}\text{sin}\theta_{\text{ng}}+g_{y}\text{cos}\theta_{\text{ng}}\right]\hat{x}^{\text{(c)}}\hat{Z}_{L}\nonumber \\
 & +\left[g_{x}\text{sin}\theta_{\text{ng}}-g_{y}\text{cos}\theta_{\text{ng}}\right]\hat{y}^{\text{(c)}}\hat{Z}_{L}\label{eq:longNG}
\end{alignat}
Unlike for the geometric bright/dark construction we note that the
last line of Eq.~\eqref{eq:longNG} does \emph{not} vanish for $\theta_{\text{ng}}\equiv\theta_{\text{tw}}$
(we recall from Eq.~\eqref{eq:long} that $g_{x}\sim\text{sin}\theta_{\text{tw}}$$/x_{\text{zpf}}$
and $g_{y}\sim\text{cos}\theta_{\text{tw}}$$/y_{\text{zpf}}$ ).
To obtain the bright/dark mode structure in this construction we must
rather set $\theta_{\text{ng}}$ such that

\begin{equation}
\text{tan}\theta_{\text{ng}}\equiv\frac{g_{x}}{g_{y}}=\sqrt{\frac{\omega_{y}}{\omega_{x}}}\text{tan}\theta_{\text{tw}}.
\end{equation}
In other words the non-geometric bright/dark modes cannot be interpreted
as the motion along/perpendicular to the cavity axis.

Using trigonometric identities we also find

\begin{equation}
\text{sin}\theta_{\text{ng}}=\frac{g_{x}}{\sqrt{g_{x}^{2}+g_{y}^{2}}},\qquad\text{cos}\theta_{\text{ng}}=\frac{g_{y}}{\sqrt{g_{x}^{2}+g_{y}^{2}}}.
\end{equation}
We can thus recast Eq.~\eqref{eq:longNG} in the form

\begin{alignat}{1}
\frac{\hat{H}}{\hbar} & =\frac{1}{4}\omega_{b}\left[\hat{x}_{b}^{2}+\hat{p}_{b}^{2}\right]+\frac{1}{4}\omega_{d}\left[\hat{x}_{d}^{2}+\hat{p}_{d}^{2}\right]\nonumber \\
 & +g_{bd}\left[\hat{x}_{b}\hat{x}_{d}+\hat{p}_{b}\hat{p}_{d}\right]+g_{b}\hat{x}_{b}\hat{Z}_{L},\label{eq:90}
\end{alignat}
where the frequencies are given by

\begin{alignat}{1}
\omega_{b}^{2} & =\frac{g_{x}^{2}\omega_{x}+g_{y}^{2}\omega_{y}}{g_{x}^{2}+g_{y}^{2}},\\
\omega_{d}^{2} & =\frac{g_{x}^{2}\omega_{y}+g_{y}^{2}\omega_{x}}{g_{x}^{2}+g_{y}^{2}},
\end{alignat}
and the couplings reduce to

\begin{alignat}{1}
g_{b} & =\sqrt{g_{x}^{2}+g_{y}^{2}},\label{eq:gbNG}\\
g_{bd} & =\frac{(\omega_{y}-\omega_{x})g_{x}g_{y}}{2\sqrt{g_{x}^{2}+g_{y}^{2}}}.\label{eq:gbdNG}
\end{alignat}

\subsubsection{Comparison with the geometric case}

The note that the geometric (Eqs.~\eqref{eq:rotationXXC} and \eqref{eq:rotationPPC})
and non-geometric transformation (Eqs.~\eqref{eq:rotationC1} and
\eqref{eq:rotationC2}) approximately match if the frequency differences
are not too large. Let us see how to explicitly recover Eqs.~\eqref{eq:gbNG}
and \eqref{eq:gbdNG} in the limit $\omega_{x}\sim\omega_{y}$ from
Eqs.~\eqref{eq:rotationXXC} and \eqref{eq:rotationPPC}. We start
from the last line of Eq.~\eqref{eq:long} and using the fact that
it vanishes we can express the angle as:

\begin{equation}
\text{tan}(\theta_{\text{tw}})=\sqrt{\frac{\omega_{x}}{\omega_{y}}}\frac{g_{x}}{g_{y}}.\label{eq:99}
\end{equation}
Using trigonometric formulae we also find:

\begin{alignat}{1}
\text{sin}(\theta_{\text{tw}}) & =\frac{\sqrt{\omega_{x}}g_{x}}{\sqrt{\omega_{x}g_{x}^{2}+\omega_{y}g_{y}^{2}}},\label{eq:100}\\
\text{cos}(\theta_{\text{tw}}) & =\frac{\sqrt{\omega_{y}}g_{y}}{\sqrt{\omega_{x}g_{x}^{2}+\omega_{y}g_{y}^{2}}}.\label{eq:101}
\end{alignat}
We now insert Eqs.~\eqref{eq:100} and \eqref{eq:101} in Eqs.~\eqref{eq:gbdef}
and \eqref{eq:gbddef} (where we use the explicit expressions for
the frequencies in Eqs.~\eqref{eq:omegab} and \eqref{eq:omegad}).
We now finally write $\omega_{x}=\omega_{y}+\delta\omega$ and Taylor
expand the couplings in $\delta\omega$: 
\begin{alignat}{1}
g_{bd}\approx & \frac{g_{x}g_{y}(\omega_{y}-\omega_{x})}{(g_{x}^{2}+g_{y}^{2})}+\mathcal{O}(\delta\omega^{2}),\label{eq:gbddef2}\\
g_{b}\approx & \sqrt{g_{x}^{2}+g_{y}^{2}}+\frac{(g_{y}^{2}-g_{x}^{2})(\omega_{y}-\omega_{x})}{4\text{\ensuremath{\omega_{x}}}\sqrt{g_{x}^{2}+g_{y}^{2}}}+\mathcal{O}(\delta\omega^{2}).\label{eq:gbdef2}
\end{alignat}
The couplings arising from the non-geometric transformation can be
thus seen as a limiting case of the couplings induced by the geometric
transformation (compare with in Eqs.~\eqref{eq:gbddef} and \eqref{eq:gbdef}).
There are however important differences. First, we note that the expression
of $g_{bd}$ in Eq.~\eqref{eq:gbdNG} is larger by a factor $2$
with respect to the coupling obtained in Eq.~\eqref{eq:gbddef2}.
Loosely speaking, this difference arises as the coupling $g_{bd}$
in Eq.~\eqref{eq:90} couples both $\hat{x}_{b},\hat{x}_{d}$ and
$\hat{p}_{b},\hat{p}_{d}$ (i.e. two terms using the non-geometric
construction) while in Eq.~\eqref{eq:bdh} it couples only $\hat{x}_{b},\hat{x}_{d}$
( a single term in the geometric construction). Second, we note that
$g_{d}$ in Eq.~\eqref{eq:gbNG} contains only the lowest order term
of the expansion in Eq.~\eqref{eq:gbdef2} which contains also a
contribution $\sim\delta\omega$.

\subsection{2D cooling formulae for bright/dark modes\label{sec:2D-cooling-formulae}}

In this section we derive the 2D optomechanical cooling rates for
the dark/bright mode. We start from the Hamiltonian in Eq.~(\ref{eq:HamiltonianBD})
and write Hamilton's equations of motion:

\begin{alignat}{1}
\dot{\hat{x}}_{b} & =\omega_{b}\hat{p}_{b},\label{eq:xbdot}\\
\dot{\hat{p}}_{b} & =-\omega_{b}\hat{x}_{b}-2g_{bd}\hat{x}_{d}-2g_{b}\hat{Z}_{L},\\
\dot{\hat{x}}_{d} & =\omega_{d}\hat{p}_{d},\label{eq:xddot}\\
\dot{\hat{p}}_{d} & =-\omega_{d}\hat{x}_{d}-2g_{db}\hat{x}_{b},\\
\dot{\hat{Z}}_{L} & =-\Delta\hat{P}_{L},\\
\dot{\hat{P}}_{L} & =\Delta\hat{Z}_{L}-2g_{b}\hat{x}_{b}.\label{eq:Pdot}
\end{alignat}
where we have introduced $g_{db}=g_{bd}$ to ease the reading of the
equations.

In the following we will consider also non-conservative terms (damping
and input noise) which we have previously omitted for clarity of presentation.
We transform Eqs.~(\ref{eq:xddot})-(\ref{eq:Pdot}) to second order
differential equations by eliminating the momenta, and express the
resulting equations in Fourier space:

\begin{alignat}{1}
\hat{x}_{b}(\omega)= & J_{bd}(\omega)\hat{x}_{d}(\omega)+J_{bY}(\omega)\hat{Z}_{L}(\omega)+\tilde{x}_{b,\text{in}}(\omega),\label{eq:xb}\\
\hat{x}_{d}(\omega)= & J_{db}(\omega)\hat{x}_{b}(\omega)+\tilde{x}_{d,\text{in}}(\omega),\label{eq:xd}\\
\hat{Z}_{L}(\omega)= & J_{Yb}(\omega)\hat{x}_{b}(\omega)+\tilde{Z}_{L,\text{in}}(\omega),\label{eq:Y}
\end{alignat}
which can be readily solved. The solutions will be labelled as $\hat{x}_{b}^{\text{2D}}(\omega)$,
$\hat{x}_{d}^{\text{2D}}(\omega)$ and $\hat{Z}_{L}(\omega)$ (each
solution in general depends on all three input noises $\tilde{x}_{b,\text{in}}(\omega)$,
$\tilde{x}_{d,\text{in}}(\omega)$ and $\tilde{Z}_{L,\text{in}}(\omega)$).
The frequency dependent coupling coefficients are given by

\begin{alignat}{1}
J_{bd}(\omega) & =-2g_{bd}\chi_{b}(\omega),\label{eq:Jbd}\\
J_{db}(\omega) & =-2g_{db}\chi_{d}(\omega),\label{eq:Jdb}\\
J_{bY}(\omega) & =-2g_{b}\chi_{b}(\omega),\label{eq:JbY}\\
J_{Yb}(\omega) & =-ig_{b}\eta(\omega),\label{eq:JYb}
\end{alignat}
where the susceptibilities are given by

\begin{alignat}{1}
\chi_{b,d}(\omega) & =\frac{\omega_{b,d}}{-\omega^{2}+\omega_{b,d}^{2}-i\omega_{b,d}\gamma},\\
\eta(\omega) & =\frac{1}{-i(\omega+\Delta)+\frac{\kappa}{2}}-\frac{1}{i(-\omega+\Delta)+\frac{\kappa}{2}}.
\end{alignat}

To find the self-energy for the bright mode, $\hat{x}_{b}$, we need
to solve (\ref{eq:xd}) and (\ref{eq:Y}) for $\hat{x}_{d}\equiv\hat{x}_{d}(\hat{x}_{b})$
and $\hat{Z}_{L}\equiv\hat{Z}_{L}(\hat{x}_{b})$ and insert the expression
in Eq.~(\ref{eq:xb}) for the bright mode. To find the self-energy
for the dark mode, $\hat{x}_{d}$, we proceed in a completely analogous
way -- we solve (\ref{eq:xb}) and (\ref{eq:Y}) for $\hat{x}_{b}\equiv\hat{x}_{b}(\hat{x}_{d})$
and $\hat{Z}_{L}\equiv\hat{Z}_{L}(\hat{x}_{d})$ and insert the expression
in Eq.~(\ref{eq:xd}) for the bright mode. From the imaginary parts
of the self-energies we can then readily extract the optomechanical
cooling rates:

\begin{alignat}{1}
\Gamma_{\text{opt,b}} & \equiv\text{Im}\left[\frac{J_{bd}(\omega_{b})J_{db}(\omega_{b})+J_{bY}(\omega_{b})J_{Yb}(\omega_{b})}{\chi_{b}(\omega_{b})}\right],\\
\Gamma_{\text{opt,d}} & \equiv\text{Im}\left[\frac{1}{\chi_{d}(\omega_{d})}\left[\frac{J_{db}(\omega_{d})J_{bd}(\omega_{d})}{1-J_{bY}(\omega_{d})J_{Yb}(\omega_{d})}\right]\right].
\end{alignat}
We now consider the ideal case $\theta=\pi/4$ where $\omega_{bd}\equiv\omega_{b}=\omega_{d}$
(see Appendix~\ref{subsec:pi/4}) and set the detuning to $\Delta=-\omega_{bd}$.
After some algebra we eventually find

\begin{alignat}{1}
\Gamma_{\text{opt,b}} & \equiv\text{Im}\left[2ig_{b}^{2}\eta(\omega_{bd})+4g_{bd}^{2}\chi_{d}(\omega_{bd})\right]\approx\text{\ensuremath{\frac{4g_{b}^{2}}{\kappa}}},\\
\Gamma_{\text{opt,d}} & \equiv\text{Im}\left[\frac{4g_{bd}^{2}\chi_{b}(\omega_{bd})}{1-2ig_{b}^{2}\chi_{b}(\omega_{bd})\eta(\omega_{bd})}\right]\approx\frac{g_{bd}^{2}\kappa}{g_{b}^{2}}.
\end{alignat}
The corresponding phonon occupancies can be roughly estimated as $n_{j}\sim\Gamma/\Gamma_{\text{opt,j}}$,
where $\Gamma$ is the motional heating rate and $j=b,d$ (as such
a procedure gives only a crude estimate for the phonon occupancy we
assume the motional heating rate is approximately equal for the bright/dark
mode).

To obtain a quantitative estimate for the phonon occupancies one has
to consider the power spectral densities (PSDs): $S_{x_{b}x_{b}}(\omega)=\langle|\hat{x}_{b}^{\text{2D}}(\omega)|^{2}\rangle$
and $S_{x_{d}x_{d}}(\omega)=\langle|\hat{x}_{d}^{\text{2D}}(\omega)|^{2}\rangle$.
These are related to the corresponding phonon occupancies as the area
under the PSD curve~\citep{bowen2015quantum}: 
\begin{equation}
n_{j}=\frac{1}{2\pi}\int_{-\infty}^{\infty}S_{x_{j}x_{j}}(\omega)d\omega-\frac{1}{2},\label{Phon-1}
\end{equation}
where $j=b,d$. We finally note that $n_{b}$ can be reliably extracted
from the optical (heterodyne) PSD, but extracting $n_{d}$ is non-trivial
and requires an indirect inference method (see the last paragraph
of Sec.~\ref{subsec:Understanding-2D-cooling}).

\bibliographystyle{unsrt}
\bibliography{2Dbiblio}

\end{document}